\documentclass{article}

\usepackage{jheppub}
\usepackage[normalem]{ulem}

\usepackage{amsmath, amssymb, amsthm} 
\usepackage{mathtools}
\usepackage{thmtools}
\usepackage{bm}
\usepackage{dsfont}
\usepackage{braket}
\usepackage{graphicx}
\usepackage{parskip}

\numberwithin{equation}{section}

\usepackage{enumitem}
\setlist[itemize]{noitemsep}
\setlist[description]{noitemsep}

\allowdisplaybreaks

\usepackage[dvipsnames]{xcolor}
\usepackage{tikz}
\usetikzlibrary{positioning, tikzmark}

\usepackage[numbers, sort&compress]{natbib}
\usepackage{hyperref}
\hypersetup{
    linkcolor=[rgb]{0.91, 0.24, 0.43}, 
    citecolor=[rgb]{0.28, 0.24, 0.75}
    }

\newcommand{\dd}{\text{d}}
\newcommand{\DD}{\mathrm{D}}
\newcommand{\pp}{\partial}
\newcommand{\PP}{\mathrm{P}}
\newcommand{\ii}{\mathrm{i}}
\newcommand{\ee}{\mathrm{e}}

\newcommand{\ul}{\underline}
\newcommand{\ol}{\overline}

\newcommand{\zb}{\overline{z}}
\renewcommand{\[}{\left[}
\renewcommand{\]}{\right]}
\renewcommand{\(}{\left(}
\renewcommand{\)}{\right)}

\renewcommand{\epsilon}{\varepsilon}
\newcommand{\dzz}{\dd z \, \dd\zb \,}

\newtheorem{identity}{Identity}

\usepackage{multirow}

\usepackage{tabularray} 
\UseTblrLibrary{booktabs}

\begin{document}

\date{\today}

\title{Holomorphic fermions in $\text{AdS}_{2}$}

\author[a,b,c]{Calvin Y.-R. Chen}
\author[c]{and Lukas W. Lindwasser}

\affiliation[a]{Max Planck--IAS--NTU Center for Particle Physics, Cosmology and Geometry, Taipei 10617, Taiwan}
\affiliation[b]{Leung Center for Cosmology and Particle Astrophysics, Taipei 10617, Taiwan}
\affiliation[c]{Department of Physics and Center for Theoretical Physics, National Taiwan University, Taipei 10617, Taiwan}

\emailAdd{cyrchen@ntu.edu.tw}
\emailAdd{llindwasser@ntu.edu.tw}

\abstract{On two-dimensional Anti-de Sitter space, Majorana fermions with a discrete set of finely tuned, non-zero mass parameters possess conserved spinor--tensors that are symmetric and traceless.
In complex coordinates, these correspond to pairs of holomorphic and anti-holomorphic half-integer spin currents and admit Laurent series mode expansions reminiscent of standard two-dimensional conformal field theories. 
We exploit this fact in order to fully parameterise the classical phase space associated with the equation of motion in terms of arbitrary (anti-)holomorphic functions. 
We further show that this parameterisation can be written covariantly in terms of a conformal Killing spinor--tensor, hinting at a hidden conformal symmetry. 
We also show they possess chiral algebras which are strict subalgebras of the one present in the corresponding theory on flat space.
}

\maketitle

\newpage 

\section{Introduction}

In two-dimensional ($2d$) quantum field theories, the existence of (anti-)holomorphic currents is exceptional, as they are associated with an infinite-dimensional symmetry algebra which severely constrains the theories in question. 
Physically, in Lorentzian signature, this (anti-)holomorphicity translates to current density profiles which travel along local lightcones. 
These are preserved under conformal transformations, suggesting that the dynamics of quantum field theories generated by such currents are conformally invariant.
One would therefore expect such currents to be a feature of massless theories. 
Recent work has shown, however, that they also arise in certain massive theories on spacetimes with non-zero constant Ricci curvature $R$.

Let us consider $2d$ Anti-de Sitter (AdS) space --- in complex coordinates, its invariant interval takes the isothermal form
\begin{equation}
    \dd s^2=\Omega(z,\zb)^2\dzz,
    \label{eq: ads metric complex}
\end{equation}
with the conformal factor
\begin{equation}
    \Omega(z,\zb)^2 = \frac{-4L^{2}}{(z-\zb)^{2}},
    \label{eq: ads conformal factor}
\end{equation}
so that $R=-2/L^{2}$. 
On this, it was found that there are non-conformally coupled scalar fields that possess a pair of such holomorphic and anti-holomorphic currents, depending only on $z$ and $\zb$ \cite{Farnsworth:2024yeh}. 
These scalar fields occur at discrete mass parameter values labelled by integers $k\in\mathbb{Z}_{\geq 0}$, with $k=0$ coinciding with the conformally coupled mass value.\footnote{In $\mathrm{dS}_2$, these special mass parameter values can be anticipated from representation theory \cite{Sun:2021thf, Schaub:2024rnl, Hinterbichler:2026xqf} --- from this perspective, the field strength is associated with their extended shift symmetries.
On the face of it, it is concerning that these mass values are tachyonic when $R>0$.
Nevertheless, these theories are associated with the discrete series representation of the de Sitter isometry group, and are rendered unitary after properly dealing with the shift symmetries.
The, perhaps more familiar, symmetry of the massless scalar field under constant shifts is a special case of this \cite{Tolley:2001gg}.} 
These are examples of a large class of theories enjoying extended shift symmetries in maximally symmetric spacetimes \cite{Bonifacio:2018zex, Hinterbichler:2022vcc}.
The non-trivial examples with $k>0$ appear in various contexts.

For instance, $k=1$ fields arise as transverse fluctuations of brane embeddings, where the shift is the linearisation of spontaneously broken bulk isometries and the non-linear completion is of Dirac--Born--Infeld (DBI) type, with the Galileon arising in its non-relativistic limit \cite{Born:1934gh, Leigh:1989jq, Luty:2003vm, Nicolis:2008in, deRham:2010eu, Hinterbichler:2014cwa,Hinterbichler:2015pqa} --- analogous constructions exist on (A)dS
\cite{Clark:2005ht, Goon:2011uw, Goon:2011qf, Burrage:2011bt}. 
In AdS$_2$ this is realised by a string worldsheet, whose transverse modes
have scaling dimension $\Delta = 2$ \cite{Metsaev:1998it, Drukker:2000ep, Giombi:2017cqn, Beccaria:2026ffm, Qiao:2026ijh,Gabai:2026myo}. 
In flat space such theories are characterised by enhanced soft limits \cite{Cheung:2014dqa,
Cheung:2016drk, Roest:2019oiw, Roest:2019dxy}.
These theories have also been studied through the representation theory of the de Sitter group \cite{Joung:2007je, RiosFukelman:2023mgq, Anninos:2023lin,Hinterbichler:2026xqf} and the quantisation of scalar fields at discrete tachyonic masses \cite{Allen:1985ux, Bros:2010wa, Epstein:2014jaa}.
Scalar fields with these special masses also arise in the free sector of flux compactifications in string theory and M-theory exhibiting scale separation \cite{Aharony:2008wz, Conlon:2021cjk, Apers:2022vfp, Apers:2022zjx, Apers:2022tfm, Ning:2022zqx, Andriot:2023fss, VanHemelryck:2025qok, Farakos:2025bwf, Miao:2025rgf, Tringas:2025bwe}, linearised variations of entanglement entropy in conformal field theory (CFT) \cite{deBoer:2015kda}, and certain linear perturbations to near-horizon geometries of extremal black holes and branes \cite{Horowitz:2023xyl, Chen:2024sgx, Chen:2025sim}. 

The holomorphic and anti-holomorphic currents are the two independent components of a conserved, symmetric, and traceless rank-$k+1$ tensor.
In \cite{Chen:2026kjo}, we used their (anti-)holomorphicity to fully parameterise the classical phase space of these scalars, identify large classes of symmetry transformations, and characterise the associated chiral algebras.
Despite not being conformally coupled, these scalar fields admit an equivalent description in terms of a symmetric and traceless rank-$k$ conformal Killing tensor, whose equations of motion are covariant with respect to conformal transformations \cite{Chen:2026kjo,Lindwasser:2026exa}. 
In $\mathrm{AdS}_2$ and dS$_{2}$, this was respectively established for conformal Dirichlet and periodic boundary conditions. 
This is closely related to the observation in \cite{Farnsworth:2024yeh} that all physical correlation functions obey conformal Ward identities in $\mathrm{dS}_2$.

\hspace{2em} In this manuscript, we will extend the classification of (anti-)holomorphic currents in $\mathrm{AdS}_2$ initiated in \cite{Chen:2026kjo} to fermionic theories, relying on the discussions in \cite{Bonifacio:2023prb, Letsios:2025pqo}.
Given the AdS zweibein $e$, we take the action for a mass $m$ Dirac field $\Psi$ in Euclidean signature to be
\begin{equation}
    S = \int \dzz \, e\, \ol{\Psi} \,(\gamma^{\mu}\DD_{\mu}-\ii m\gamma_*)\,\Psi,
     \label{eq: dirac action}
\end{equation}
where $\DD$ is the covariant derivative and $\gamma_*$ is the chirality operator.
In \cite{Bonifacio:2023prb}, it was shown that on $\mathrm{(A)dS}$ spacetimes, this theory admits a field strength $\mathcal{F}$ that is a conserved, symmetric, and traceless rank $k+1$ spinor--tensor when the mass is tuned to $m=m_{k}$, with
\begin{equation}
	m_{k} L \equiv k+1,\quad k \in \mathbb{Z}_{\geq 0}.
	\label{eq: discrete series masses}
\end{equation}
In components, this field strength is 
\begin{equation}
     \mathcal{F}_{\mu_1\cdots\mu_{k+1}}=\DD^{+}_{(\mu_1}\cdots\DD^{+}_{\mu_{k+1})_T}\Psi,\quad \DD^{+}_\mu=\DD_\mu +\frac{\ii}{2L}\gamma_*\gamma_\mu,
     \label{eq: field strength}
\end{equation}
where $(\cdots)_T$ denotes the totally symmetric and traceless projection of the contained indices.
This contains two independent components, which in terms of complex coordinates $(z,\zb)$, are $\mathcal{F}_{z\cdots z}$ and $\mathcal{F}_{\zb\cdots\zb}$. 
This current is covariantly conserved in the sense that
\begin{equation}
    \DD^{\lambda} \mathcal{F}_{\lambda \mu_{2} \dots \mu_{k+1}} = 0,
    \label{eq: covariant conservation}
\end{equation}
so that on-shell
\begin{equation}
    \mathcal{F}_{z\cdots z} = \begin{pmatrix}
    0 \\
    \Omega^{-1/2}F(z,\zb)
    \end{pmatrix},\quad 
    \mathcal{F}_{\zb\cdots\zb} = \begin{pmatrix}
        \Omega^{-1/2}\ol{F}(z,\zb)\\
        0
    \end{pmatrix},
\end{equation}
with
\begin{equation}
    \ol\pp F(z,\zb)=0,\quad \pp \ol{F}(z,\zb)=0.
\end{equation}
In other words: The field strength tensor splits into (anti-)holomorphic components. 
As in the case of the discrete series scalars, correlation functions of these holomorphic and anti-holomorphic currents will be identical to those seen in $2d$ CFTs.

Holomorphicity of $F(z)$ and anti-holomorphicity of $\ol{F}(\zb)$ imply that there is a local operation on $\Psi$ which isolates the (anti-)holomorphic sectors of the solution space. 
Just as in the case of the discrete series scalar, we are able to write the general solution to the Dirac equation with $m=m_{k}$
\begin{equation}
    \(\gamma^{\mu}\DD_{\mu}-\ii m_k \gamma_{*}\)\Psi = 0
	\label{eq: dirac eq}
\end{equation}
in the form
\begin{equation}
	\Psi=\DD^{-}_{\mu_1}\cdots\DD^{-}_{\mu_{k+1}}\mathcal{Y}^{\mu_1\cdots\mu_{k+1}},\quad \DD^{-}_{\mu} = \DD_\mu  -\frac{\ii}{2L}\gamma_*\gamma_\mu,
    \label{eq: fermion split}
\end{equation}
in terms of a symmetric and traceless rank-$k+1$ spinor--tensor $\mathcal{Y}^{\mu_1\cdots\mu_{k+1}}$ with 
\begin{equation}
	\mathcal{Y}^{z \cdots z}=\begin{pmatrix}
		0 \\
		\Omega^{-1/2} Y(z)
	\end{pmatrix},\quad 
	\mathcal{Y}^{\zb \cdots \zb}=\begin{pmatrix}
		\Omega^{-1/2}\tilde Y(\zb)\\
		0
	\end{pmatrix}.
    \label{eq:YYbarsoln}
\end{equation}
This is equivalent to a spinor--tensor which obeys the coupled system of equations 
\begin{equation}
    \DD_{(\mu_1}\mathcal{Y}_{\mu_2\cdots\mu_{k+2})_T}=0,\quad \gamma_{(\mu_1}\mathcal{Y}_{\mu_2\cdots\mu_{k+2})_T}=0.
    \label{eq: YCKST intro}
\end{equation}
As a coupled system, this is covariant with respect to conformal transformations, hinting at a hidden conformal symmetry on other constant curvature spacetimes. 
In that regard, we argue that it is informative to reformulate the theory in terms of these conformal Killing spinor--tensors.

\subsection*{Organisation}

Our manuscript is organised as follows.
In section \ref{sec: set-up} we will first discuss the classical theory and quantisation of a Majorana fermion at the tuned masses \eqref{eq: discrete series masses}, before moving onto complex coordinates.
Then, in section \ref{sec: holomorphic splitting}, we will demonstrate how the classical phase space of this theory can be fully parameterised in terms of two arbitrary functions, which allows us to display the hidden conformal symmetry in the theory.
We elucidate its origin in section \ref{sec:Hiddenconformalsymmetry}, by relating this to the existence of a conformal Killing spinor--tensor.
Finally, in section \ref{sec:chiralalgebra}, we will show that these theories possess a chiral algebra associated with arbitrary (anti-)holomorphic functionals of the (anti-)holomorphic current $F$ ($\ol{F}$), which is a subalgebra of the chiral algebra present in the theory of a free massless fermion on two-dimensional Minkowski space.

Some details of calculations and derivations have been relegated to the appendices.
In appendix~\ref{app: fermion conventions}, we spell out the conventions we employ for describing fermions in two-dimensional Anti-de Sitter space both in global and complex coordinates.
In appendix~\ref{app: field strength holomorphicity}, we explicitly show that the field strength tensor \eqref{eq: field strength} is conserved in complex coordinates.
Then, in appendix~\ref{app: holomorphic splitting} we present two derivations for the splitting of the classical phase space associated with the holomorphic fermions.
We show that a similar decomposition occurs for the field strength in appendix~\ref{app: field strenght splitting}.
Finally, in appendix~\ref{app:invariance of action}, we explicitly show that the action \eqref{eq: dirac action} is invariant under arbitrary chiral transformations. 

\subsection*{Conventions}

We focus on two-dimensional Anti-de Sitter space, for which we take the radius to be $L$ so that $R= -2/L^{2}$.
To describe fermions thereon, we introduce the zweibeine $e$.
This is associated with a unique Levi-Civita connection, with respect to which we denote covariant derivatives by $\DD$.
In order to distinguish indices in the coordinate basis and the local frame, we will \underline{underline} the latter.

In Lorentzian signature, we will work in mostly-plus convention, i.e. with $(\text{time}, \text{space}) = (-,+)$.
Here, we will use global coordinates $(\tau,\rho)$.
We analytically continue to $(+,+)$ in Euclidean signature, where we work in complex coordinates $(z,\zb)$.
It is convenient to use a shorthand to denote the (anti-)holomorphic components of the partial and covariant derivatives by $\pp \equiv \pp_{z}$ ($\ol{\pp} \equiv \pp_{\zb}$) and $\DD \equiv \DD_{z}$ ($\ol{\DD} \equiv \DD_{\zb}$) respectively.
Tensor indices are (anti-)symmetrised with unit weight and are denoted by round (square) brackets, and the subscript $T$ further denotes the traceless projection.

For convenience, we will occasionally also slightly abuse notation and denote components of an ordered tuple (such as a vector or spinor) by the set inclusion symbols $\in$ and $\ni$.

\section{Discrete series fermions \label{sec: set-up}}

The subject of interest is the Majorana fermion $\Psi$ described by the Dirac action \eqref{eq: dirac action} with $m=m_{k}$ in \eqref{eq: discrete series masses}.
Notably, its $\mathrm{dS}_2$ counterparts were previously studied in \cite{Bonifacio:2023prb, Letsios:2025pqo}.

In this section, we solve the classical equation of motion in global coordinates and use the resulting mode functions to canonically quantise the theory.
We will then switch to complex coordinates and introduce the holomorphic and anti-holomorphic components of the field strength tensor \eqref{eq: field strength}.
Conventions for fermions in both coordinates are discussed in appendix \ref{app: fermion conventions}.

\subsection{Fermions in global AdS$_{2}$}

Consider Lorentzian $\text{AdS}_{2}$ with radius $L$ and Ricci scalar $R = -2/L^{2}$. 
Here, we will use conformal coordinates $(\tau,\rho)$, which cover the conformal compactification of the global patch.
The metric takes the form
\begin{equation}
    \dd s^{2} = \frac{L^{2}}{\cos(\rho)^{2}}(-\dd\tau^{2} + \dd\rho^{2}),
    \label{eq: strip metric}
\end{equation}
where $\rho \in (-\pi/2,\pi/2)$ and $\tau \in \mathbb{R}$.

We are interested in the massive Dirac field $\Psi$ described by the action 
\begin{equation}
    S_{\text{massive}} = -\int \dd^{2} x\, e\, \ol{\Psi} \,(\gamma^{\mu}\DD_{\mu}+m)\,\Psi,
\end{equation}
where the conjugate spinor is defined as $\ol{\Psi} \equiv \ii\Psi^{\dag}\gamma^{\ul{0}}$.
We must impose Dirichlet boundary conditions, so that $\Psi(\tau,\pm\pi/2)=0$. 
As its classical equation of motion, the field $\Psi$ obeys the Dirac equation
\begin{subequations}
    \begin{align}
        \big(\gamma^{\mu}\DD_{\mu}+ m\big)\Psi & = 0 \\
        \ol{\Psi}\big(\gamma^{\mu}\overleftarrow{\DD}_\mu - m\big) &= 0.
    \end{align}
    \label{eq: dirac eq lorentzian}%
\end{subequations}
Because $2 \text{ (mod } 8)= 2$, we may and shall restrict ourselves to Majorana spinors.
Since we have chosen a real representation for the $\gamma$-matrices, the charge conjugate of $\Psi$ is just $\Psi^{*}$, and we may write
\begin{equation}
    \Psi = 
    \begin{pmatrix}
        \psi_{+} \\
        \psi_{-}
    \end{pmatrix},
    \label{eq: global components}
\end{equation}
where $\psi_{\pm}$ are its real Majorana--Weyl components.
More details on conventions for fermions are presented in appendix \ref{sapp: global conventions}.

\subsection{Quantisation of the discrete series}

We are interested in fermions with the discrete set of masses defined by \eqref{eq: discrete series masses}.
As was shown in \cite{Bonifacio:2023prb}, these are related to the discrete series representation on dS$_{2}$ and possess extended shift symmetries.
Instead of working with the mass $m$ directly, it is sometimes convenient to define the $\text{AdS}_{2}$ scaling dimension
\begin{equation}
    \Delta = mL+1/2,
\end{equation}
which takes half-integer values $\Delta=k+3/2$ when $m=m_{k}$.

Let us now briefly review properties of the solutions to \eqref{eq: dirac eq lorentzian}, following \cite{Blanco:2022quy} specified to half-integer $\Delta$.
First note that the space of these solutions is equipped with an invariant inner product, which is the Dirac inner product:
\begin{equation}
    \begin{aligned}
        \( \Psi_{1},\Psi_{2}\) &= \int_{-\pi/2}^{\pi/2} \frac{\dd \rho}{\cos(\rho)} \Psi_{1}(\tau,\rho)^{\dag} \Psi_{2}(\tau,\rho) \\
        &= \int_{-\pi/2}^{\pi/2} \frac{\dd \rho}{\cos(\rho)} \(\psi_{1,-}^{*}\psi_{2,-} + \psi_{1,+}^{*}\psi_{2,+}\).
    \end{aligned}
    \label{eq: dirac inner product}
\end{equation}
This defines a Hermitian inner product.
To define the quantised theory, we seek a decomposition of the classical phase space of solutions to the Dirac equation \eqref{eq: dirac eq lorentzian} which is orthogonal with respect to this.

To that end, we will now construct a set of solutions which are normalisable with respect to the Dirac inner product.
Due to the presence of the global timelike Killing vector $\pp/\pp\tau$, it is sensible to decompose our space of solutions in terms of eigenmodes of its Lie derivative:
\begin{equation}
    \psi_{\omega}(\tau,\rho) =\cos(\rho)^{1/2} f_{\omega}(\rho) \ee^{-\ii\omega\tau},\quad \omega > 0.
\end{equation}
Plugging this into the Dirac equation \eqref{eq: dirac eq lorentzian}, we see that the spatial parts $f_\omega$ ought to satisfy
\begin{equation}
    \(\omega -\ii \sigma_{3}\pp_{\rho} + \frac{\Delta-1/2}{\cos(\rho)}\sigma_{2}\)f_{\omega}(\rho) = 0.
\end{equation}
The solutions to this which have positive energy, are square-integrable, and $C^{1}$ on $\rho\in(-\pi/2, \pi/2)$ are subject to the quantisation condition $\omega \in \mathbb{Z} + \frac{1}{2}$ and $\omega > k+1$.
Explicitly, we pick
\begin{equation}
    f_\omega(\rho)=N_\omega\cos(\rho)^{k+1}\begin{pmatrix}
        A_\omega(\rho)+\ii B_{\omega}(\rho)\\
        A_\omega(\rho)-\ii B_{\omega}(\rho)
    \end{pmatrix},\quad 
\end{equation}
where
\begin{equation}
    A_\omega(\rho)=\sqrt{1+\sin(\rho)}P_{\omega-k-3/2}^{(k+1/2,k+3/2)}(\sin(\rho)), \quad         B_\omega(\rho)=\sqrt{1-\sin(\rho)}P_{\omega-k-3/2}^{(k+3/2,k+1/2)}(\sin(\rho)).
\end{equation}
Further, the normalisation constants are chosen to be
\begin{equation}
    N_\omega=\frac{\sqrt{\Gamma(\omega-k-1/2)\Gamma(\omega+k+3/2)}}{2^{k+2}\Gamma(\omega)},
\end{equation}
such that $\psi_{\omega}$ are orthonormal with respect to the Dirac inner product: 
\begin{equation}
	(\psi_{\omega},\psi_{\omega'}) = \delta_{\omega,\omega'}.
\end{equation}

We can now construct the full mode expansion.
In stationary spacetimes, the orthogonal complement of the space of positive-frequency solutions is obtained by charge conjugation --- these are precisely the negative-frequency solutions $\psi^{*}$.
The full mode expansion for the solutions to the Dirac equation \eqref{eq: dirac eq lorentzian} for a massive Majorana fermion in $\text{AdS}_{2}$ in conformal coordinates hence takes the form
\begin{equation}
    \begin{aligned}
        \Psi &= \sum_{\omega \geq k+3/2} a_{\omega} \psi_{\omega} + a_{\omega}^{\dag} \psi_{\omega}^{*} \\
        &= \sum_{\omega \geq k+3/2} a_{\omega} \cos(\rho)^{1/2}f_{\omega}\ee^{-\ii\omega\tau} + a_{\omega}^{\dag} \cos(\rho)^{1/2}f_{\omega}^{*} \ee^{\ii\omega\tau},
    \end{aligned}
    \label{eq: global ads mode expansion}
\end{equation}
where we have defined the operators
\begin{equation}
    a_{\omega} = (\psi_{\omega},\Psi),\quad a_{\omega}^{\dag} = (\psi_{\omega}^{*},\Psi).
\end{equation}
These obey the anti-commutation relations
\begin{equation}
    \{a_{\omega},a^{\dag}_{\omega'}\}  = \delta_{\omega,\omega'}.
\end{equation}
There is a unique vacuum state $|0\rangle$ defined by the condition
\begin{equation}
    a_\omega|0 \rangle=0,
    \label{eq: vacuum def}
\end{equation}
and we can build our Fock space by repeatedly applying the creation operators $a_{\omega}^{\dag}$.

\subsection{Complex coordinates}

We now switch to describing the massive holomorphic fermions in Euclidean $\text{AdS}_2$ using complex coordinates on the upper half-plane (UHP).
More details on conventions for fermions in Euclidean signature complex coordinates are given in appendix \ref{sapp: UHP conventions}.

We start by analytically continuing to Euclidean time $\tau = -\ii\tau_{\text{E}}$ and defining
\begin{equation}
    z = \ee^{\tau_{\text{E}}-\ii(\rho-\pi/2)},
    \label{eq: complex coordinates}
\end{equation}
such that
\begin{equation}
    \rho = \frac{\ii}{2}\log(z/\zb) + \frac{\pi}{2},\quad \tau_{\text{E}} = \frac{1}{2}\log(z\zb).
\end{equation}
In these complex coordinates $(z,\zb)$, the metric \eqref{eq: strip metric} is brought into isothermal form \eqref{eq: ads metric complex} with conformal factor \eqref{eq: ads conformal factor}.
We choose the $\gamma$-matrices to be analytically continued in the following way
\begin{equation}
    \gamma^{\ul{0}}=\gamma_*\gamma_{\text{E}}^{\ul{2}},\quad \gamma^{\ul{1}}=\ii\gamma_*\gamma_{\text{E}}^{\ul{1}},
\end{equation}
where $\gamma_*=\sigma_3$ is the highest-weight element with $\{\gamma_*,\gamma_{\text{E}}^\mu\}=0$. 
Finally, we redefine the Dirac adjoint $\ol{\Psi}= \ol{\Psi}_{\text{E}}(-\ii \gamma_*)$, which leaves us with the Euclidean action \eqref{eq: dirac action}.
From now on, we will work in Euclidean signature unless stated otherwise, so we will drop E-subscripts.

The Majorana--Weyl components \eqref{eq: global components} of the field $\Psi$ also transform under the coordinate transformation \eqref{eq: complex coordinates}.  
We write its new components as
\begin{equation}
\label{eq:EuclPsicomp}
    \Psi=\begin{pmatrix}
        \tilde\psi\\
        \psi
    \end{pmatrix},
\end{equation}
which are related to the original components \eqref{eq: global components} via the map, 
\begin{equation}
    \psi_{+} \mapsto  \(\frac{z}{\zb}\)^{-1/4} \tilde{\psi}(z,\zb),\quad \psi_{-} \mapsto  \(\frac{z}{\zb}\)^{+1/4} \psi(z,\zb).
    \label{eq: spinor uhp mapping}
\end{equation}
In terms of these components and the $\gamma$-matrices defined in appendix \ref{sapp: UHP conventions}, the Euclidean action \eqref{eq: dirac action} becomes
\begin{equation}
    S =\int \dzz \Omega \(\tilde\psi\pp\tilde\psi+\psi\ol\pp\psi+\ii m_k\Omega\tilde\psi\psi\).
\end{equation}
In terms of $\psi$ and $\tilde{\psi}$, the equations of motion read
\begin{subequations}
	\begin{align}
        \(\ol{\pp}+ \frac{\ol{\pp}\Omega}{2\Omega}\)\psi -\ii \frac{m_k\Omega}{2} \tilde{\psi} &= 0 \\
		\(\pp + \frac{\pp\Omega}{2\Omega}\)\tilde{\psi} + \ii \frac{m_k\Omega}{2} \psi &= 0.	
	\end{align}
	\label{eq: dirac eq complex comp}%
\end{subequations}

\subsection{Field strength \label{ssec:Fieldstrength}}

The field strength tensor $\mathcal{F}$ is traceless by construction, so in complex coordinates on EAdS$_{2}$ its only independent components are $\mathcal{F}_{z\cdots z}$ and $\mathcal{F}_{\zb\cdots\zb}$. 
Let us focus on the holomorphic component for now.
Using the fact that $\gamma_{\mu}$ is covariantly constant and that $\gamma_{z}^{2} = 0 = \gamma_{\zb}^{2}$, we can write
\begin{equation}
    \mathcal{F}_{z\cdots z} = \DD^{k+1}\Psi + \ii \frac{k+1}{2L} \gamma_* \gamma_{z} \DD^{k}\Psi.
\end{equation}
Using the projection identities \eqref{eq: gamma projections} and the fact that the projectors commute with the covariant derivative, we note that the anti-holomorphic projection reads
\begin{equation}
    \begin{aligned}
        \ol{\PP} \mathcal{F}_{z\cdots z} &= \DD^{k+1}\(\ol{\PP}\Psi\) +  \ii \frac{k+1}{2L} \gamma_* \gamma_{z} \DD^{k}\Psi \\
        &= \DD^{k}\[\DD\(\ol{\PP}\Psi\) +  \ii \frac{k+1}{2L} \gamma_* \gamma_{z} \Psi\].
    \end{aligned}
    \label{eq: field strength on-shell projections}
\end{equation}
The term in brackets can be seen to vanish by acting on the Dirac equation \eqref{eq: dirac eq} with $\gamma_{z}$ and $\gamma_{\zb}$ from the left and using the projector identities \eqref{eq: gamma projections}.
In other words, this vanishes identically on-shell.
The $\mathcal{F}_{\zb\cdots\zb}$ case proceeds similarly, and we therefore have the two non-zero projections $\PP\mathcal{F}_{z\cdots z}$ and $\ol{\PP}\mathcal{F}_{\zb\cdots\zb}$.
Now, in components, covariant conservation of the field strength \eqref{eq: covariant conservation} translates into 
\begin{equation}
    \label{eq:complexconservation}
    \ol{\DD} \mathcal{F}_{z\cdots z} = 0,\quad \DD \mathcal{F}_{\zb\cdots\zb} = 0.
\end{equation}
In terms of the non-zero projections $\PP\mathcal{F}_{z\cdots z}$ and $\ol{\PP}\mathcal{F}_{\zb\cdots\zb}$, this is solved by
\begin{equation}
\label{eq:FFbardefinition}
    \PP \mathcal{F}_{z\cdots z} = \begin{pmatrix}
    0 \\
    \Omega^{-1/2}F(z)
    \end{pmatrix},\quad 
    \ol{\PP}\mathcal{F}_{\zb\cdots\zb} = \begin{pmatrix}
        \Omega^{-1/2}\ol{F}(\zb)\\
        0
    \end{pmatrix},
\end{equation}
where $F$ ($\ol{F}$) is the non-trivial on-shell component of the (anti-)holomorphic field strength tensor.
We show this explicitly in appendix \ref{app: field strength holomorphicity}.

\section{Holomorphic splitting \label{sec: holomorphic splitting}}

We now come to one of our central results.
The claim is that the phase space of the discrete series fermion \eqref{eq: dirac eq} at the mass value \eqref{eq: discrete series masses} is fully parameterised by two (anti-)holomorphic functions.
In UHP coordinates and the fermion components \eqref{eq:EuclPsicomp}, we can write this as
\begin{subequations}
    \begin{align}   
        \tilde{\psi}(z,\zb) &= \Omega^{1/2}\ol{\mathcal{D}}_{k+1}\big(\Omega^{-1}\tilde{Y}(\zb)\big) - \ii\frac{m_{k}}{2}\Omega^{-1/2}\mathcal{D}_{k}\big(\Omega \,Y(z)\big) \\
        \psi(z,\zb) &= \Omega^{1/2}\mathcal{D}_{k+1}\big(\Omega^{-1}Y(z)\big) + \ii\frac{m_{k}}{2}\Omega^{-1/2}\ol{\mathcal{D}}_{k}\big(\Omega \,\tilde{Y}(\zb)\big),
    \end{align}
    \label{eq: holomorphic splitting}%
\end{subequations}
where $Y(z)$ and $\tilde Y(\zb)$ are arbitrary holomorphic and anti-holomorphic functions.
In writing this, we have made use of the order-$k$ and -$(k+1)$ differential operators $\mathcal{D}_k$, $\ol{\mathcal{D}}_k$, $\mathcal{D}_{k+1}$, and $\ol{\mathcal{D}}_{k+1}$ defined by
\begin{equation}
	\begin{aligned}
		\mathcal{D}_{\alpha} \cdot\,&=\sum_{n=0}^{\alpha}\frac{(2\alpha-n)!}{n!(\alpha-n)!}(-1)^{\alpha-n}\frac{1}{\(z-\zb\)^{\alpha-n}}\pp^{n} \cdot \\
		&= \(\frac{1}{z-\zb}\)^{-\alpha-1}\partial^\alpha\bigg[\(\frac{1}{z-\zb}\)^{\alpha+1} \;\cdot\bigg],
	\end{aligned}
	\label{eq: mathcalD def}%
\end{equation}
and $\ol{\mathcal{D}}_\alpha$ given by the complex conjugate, introduced in \cite{Chen:2026kjo}. 
In appendix \ref{app: holomorphic splitting} we prove that \eqref{eq: holomorphic splitting} is a local solution to the equations of motion when $Y(z)$ and $\tilde{Y}(\zb)$ are arbitrary functions.
This is equivalent to the parameterisation \eqref{eq: fermion split} in the introduction, albeit in more explicit form. 

In terms of the functions $Y(z)$ and $\tilde{Y}(\zb)$, the field strength components $F(z)$ and $\ol{F}(\zb)$ defined in \eqref{eq:FFbardefinition} take particularly simple forms.
In particular, as we show in appendix \ref{app: field strenght splitting},
\begin{equation}
    F(z)=\pp^{2k+2}Y(z),\quad \ol{F}(\zb)= \ol\pp^{2k+2}\tilde{Y}(\zb).
    \label{eq: field strength parameterisation}
\end{equation}
These are manifestly (anti-)holomorphic.

\subsection{Complex mode expansions}

The parameterisation \eqref{eq: holomorphic splitting} is always a solution to the Dirac equation at the mass value \eqref{eq: discrete series masses}. What is not clear is that this mapping is surjective, i.e. that the full phase space is parameterised by \eqref{eq: holomorphic splitting}. 
To demonstrate this, it is sufficient to reproduce the mode expansion \eqref{eq: global ads mode expansion} by specifying $Y(z)$ and $\tilde{Y}(\zb)$. 
This is achieved by
\begin{subequations}
    \begin{align}
        Y(z) & = (-\ii)^{k+1} \sqrt{\frac{L}{2\pi}} \sum_{|r|>k+1/2} \frac{1}{\sqrt{\prod_{j=0}^{k}\(r^2-\(j+\frac{1}{2}\)^{2}\)}} \frac{\psi_{r}}{z^{r-k-1/2}},\\
        \tilde{Y}(\zb) &= (+\ii)^{k+1} \sqrt{\frac{L}{2\pi}} \sum_{|r|>k+1/2} \frac{1}{\sqrt{\prod_{j=0}^{k}\(r^2-\(j+\frac{1}{2}\)^{2}\)}}\frac{\psi_{r}}{\zb^{r-k-1/2}},
    \end{align}
    \label{eq: UHP Y mode expansion}
\end{subequations}
where $r\in\mathbb{Z}+1/2$, and we have introduced the notation 
\begin{equation}
    \psi_{r} \equiv a_{r},\quad \psi_{-r} \equiv a_{r}^{\dag}
\end{equation}
for $r > k+1/2$. 
These newly defined oscillator operators $\psi_r$ satisfy the anti-commutation relations
\begin{equation}
    \{\psi_r,\psi_s\}=\delta_{r+s,0}.
\end{equation}
In AdS$_{2}$, the unique boundary condition ($\Psi\to 0$ at the conformal boundary \cite{Blanco:2022quy}) implies that the general solution to the Dirac equation is completely characterised by \eqref{eq: holomorphic splitting} and \eqref{eq: UHP Y mode expansion}.

Using \eqref{eq: field strength parameterisation}, the oscillator mode expansions for the components $F(z)$ and $\ol F(\zb)$ are thus
\begin{subequations}
    \begin{align}
        F(z)&=(-\ii)^{k+1} \sqrt{\frac{L}{2\pi}} \sum_{|r|>k+1/2} \sqrt{\textstyle\prod_{j=0}^{k}\(r^2-\(j+\frac{1}{2}\)^{2}\)}\frac{\psi_{r}}{z^{r+k+3/2}} \\
        \ol{F}(\zb)&=(+\ii)^{k+1} \sqrt{\frac{L}{2\pi}} \sum_{|r|>k+1/2} \sqrt{\textstyle\prod_{j=0}^{k}\(r^2-\(j+\frac{1}{2}\)^{2}\)}\frac{\psi_{r}}{\zb^{r+k+3/2}}.
    \end{align}
    \label{eq: UHP F mode expansion}%
\end{subequations}
Up to the overall factor, this reduces to the flat space expansion $\psi(z)=\sum_r\psi_r z^{-r-1/2}$ for $k=-1$.
Note that the mode expansions \eqref{eq: UHP Y mode expansion} and \eqref{eq: UHP F mode expansion} make explicit that the following conditions are satisfied on the conformal boundary $z=\bar z$ of the UHP:
\begin{equation}
   Y(z)\sim (-1)^{k+1}\tilde{Y}(\zb)\quad (\text{as } z\to\bar z),\quad
   F(z)\sim (-1)^{k+1}\ol{F}(\zb)\quad (\text{as } z\to\bar z).
\end{equation}
The (anti-)holomorphic sectors are therefore related, as is typical for fields with Dirichlet boundary conditions.

\subsection{Correlation functions \label{ssec:correlationfunctions}}

In $\mathrm{AdS}_2$ there is a unique choice of vacuum \eqref{eq: vacuum def} for $\Delta\geq 1$ (in particular for all $m_k$), and therefore a unique two-point correlation function for the fermion $\Psi$. 
This two-point correlation function is well known, and can be found in various standard treatments, including \cite{Beccaria:2019dju,Beccaria:2026ffm}. 
For our purposes, we present the two-point correlation functions involving the field strength components $F$ and $\ol{F}$. 
To compute this explicitly, one possible strategy is to follow \cite{Farnsworth:2024yeh} by first determining the two-point correlation function for the fermion $\Psi$, and then act thereon with the appropriate differential operators. 
Instead, with the oscillator mode expansions \eqref{eq: UHP F mode expansion} in hand, we can readily compute correlation functions of $F$ and $\ol{F}$ in the vacuum directly using the anti-commutation relations the oscillator modes satisfy.
In particular, one finds
\begin{subequations}
\label{eq:F2pnt}
    \begin{align}
        \braket{F(z)F(z')} &= (-1)^{k+1}(2k+2)!\frac{L}{2\pi} \frac{1}{(z-z')^{2k+3}} \\
        \braket{F(z)\ol{F}(\ol{z}')} &= (2k+2)! \frac{L}{2\pi} \frac{1}{(z-\ol{z}')^{2k+3}}\\
        \braket{\ol{F}(\zb)\ol{F}(\ol{z}')} &= (-1)^{k+1}(2k+2)! \frac{L}{2\pi} \frac{1}{(\zb-\ol{z}')^{2k+3}}.
    \end{align}
\end{subequations}
These are identical to two-point correlation functions for global conformal primary operators with (anti-)holomorphic weights $(h_F,\ol{h}_F) = (k+3/2,0)$ and $(h_{\ol{F}},\ol{h}_{\ol{F}}) = (0,k+3/2)$ that are typically seen in $2d$ CFTs with Dirichlet boundary conditions, see e.g. \cite{Blumenhagen:2013fgp}.
Since this is a free theory, all higher-point correlation functions are fixed by Wick's theorem.
As a result, we find that the set of correlation functions generated by the operators $F$ and $\ol{F}$ is indistinguishable from those in a CFT.

\section{Hidden conformal symmetry \label{sec:Hiddenconformalsymmetry}}

As we saw in $\mathrm{AdS}_2$, the two-point correlation functions \eqref{eq:F2pnt} bear striking resemblance to two-point correlation functions in $2d$ CFTs with boundary. 
In that case however, this is not sufficient to claim that these two-point correlation functions obey \textit{conformal} Ward identities, thereby demonstrating a conformal symmetry.
Indeed, the form of the functions \eqref{eq:F2pnt} is completely determined up to normalisations by the $\mathrm{AdS}_2$ isometries \eqref{eq:AdSisometries} together with the conservation condition \eqref{eq:complexconservation}.
In this section, we will describe a concrete sense in which the relevant space of solutions to the Dirac equation at the mass value \eqref{eq: discrete series masses}, ignoring global considerations, \textit{is} invariant under arbitrary local conformal transformations.

\subsection{Conformal Killing spinor--tensor}

To that end, we need to take a step back from AdS$_{2}$ and consider a more generic $2d$ spacetime with constant Ricci scalar $R$.

Consider the Dirac equation \eqref{eq: dirac eq}.
For $R\neq 0$, the mass sequence \eqref{eq: discrete series masses} generalises to
\begin{equation}
	m_k=-\ii(k+1)\sqrt{\frac{R}{2}},\qquad k\in\mathbb{Z}_{\geq 0}.
	\label{eq: general discrete series masses}
\end{equation}
When $R<0$, $m_k\in \mathbb{R}$ and we may choose a branch such that $m_k>0$ without loss of generality. 
When $R>0$, the mass is imaginary and therefore describes a tachyonic field. 
In either case, we make the following ansatz for $\Psi$ in terms of a symmetric and traceless rank-$(k+1)$ spinor--tensor $\mathcal{Y}^{\mu_1\cdots\mu_{k+1}}$:
\begin{equation}
    \Psi=\DD^{-}_{\mu_1}\cdots\DD^{-}_{\mu_{k+1}}\mathcal{Y}^{\mu_1\cdots\mu_{k+1}},\quad \DD^{-}_\mu \cdot \equiv \DD_\mu \cdot -\sqrt{\frac{R}{8}}\gamma_*\gamma_\mu \cdot.
    \label{eq: general fermion split}
\end{equation}
As we show in appendix \ref{app: holomorphic splitting}, \eqref{eq: general fermion split} is a solution to the Dirac equation \eqref{eq: dirac eq} with $m=m_{k}$ defined by \eqref{eq: general discrete series masses} when $\mathcal{Y}^{\mu_1\cdots\mu_{k+1}}$ satisfies the conformal Killing tensor equation together with the algebraic condition:
\begin{equation}
    \DD_{(\mu_1}\mathcal{Y}_{\mu_2\cdots\mu_{k+2})_T}=0,\quad \gamma_{(\mu_1}\mathcal{Y}_{\mu_2\cdots\mu_{k+2})_T}=0.
    \label{eq: YCKST}
\end{equation}
This follows directly from the intertwining relation \eqref{eq: Y main identity} between the Dirac equation and the two equations \eqref{eq: YCKST}, proven in appendix \ref{app: holomorphic splitting}.
The intertwining relation ensures that the ansatz \eqref{eq: general fermion split} solves the Dirac equation whenever $\mathcal{Y}^{\mu_1\cdots\mu_{k+1}}$ satisfies \eqref{eq: YCKST}.
In $\mathrm{AdS}_2$ the general solution to \eqref{eq: YCKST} is \eqref{eq:YYbarsoln}.
Inserting this into the ansatz \eqref{eq: general fermion split} reproduces the holomorphic splitting in component form \eqref{eq: holomorphic splitting}. 
Note that the identity \eqref{eq: Y main identity} only guarantees that the ansatz yields solutions to the Dirac equation; it does not by itself establish that every solution is of this form. 
We will not address this question at the level of local solutions. 
Instead, we use the fact that the global structure of the spacetime and boundary conditions imposed on $\Psi$ restrict the phase space. 
In $\mathrm{AdS}_2$, the mode expansion \eqref{eq: UHP Y mode expansion} reproduces the full set of normalisable solutions \eqref{eq: global ads mode expansion}, so that \eqref{eq: general fermion split} subject to \eqref{eq: YCKST} does furnish the general solution there.

The coupled system \eqref{eq: YCKST} has special properties under Weyl rescaling. 
If a spacetime metric $g$ admits a spinor--tensor $\mathcal{Y}_{\mu_1\cdots\mu_{k+1}}$ satisfying \eqref{eq: YCKST}, a conformally related spacetime $g'=e^{2\sigma}g$ admits a spinor--tensor 
\begin{equation}
    \label{eq:hiddenconformalsymmetry}
    \mathcal{Y}'^{\mu_1\cdots\mu_{k+1}}(x')=e^{-\sigma/2}\mathcal{Y}^{\mu_1\cdots\mu_{k+1}}(x).
\end{equation}
Indeed, the coupled system is covariant with respect to this transformation, 
\begin{subequations}
    \begin{align}
        &\DD'_{(\mu_1}\mathcal{Y}'_{\mu_2\cdots\mu_{k+2})_{T'}}=e^{(2k+3/2)\sigma}\(\DD_{(\mu_1}\mathcal{Y}_{\mu_2\cdots\mu_{k+2})_T}-\frac{1}{2}\gamma^\nu\partial_\nu\sigma \gamma_{(\mu_1}\mathcal{Y}_{\mu_2\cdots\mu_{k+2})_T}\),\\
        &\gamma'_{(\mu_1} \mathcal{Y}'_{\mu_2\cdots\mu_{k+2})_{T'}}=e^{(2k+5/2)\sigma}\gamma_{(\mu_1}\mathcal{Y}_{\mu_2\cdots\mu_{k+2})_T}.
    \end{align}
\end{subequations}
This covariance on the space of solutions suggests a natural action of conformal transformations on $\Psi$ in terms of the conformal properties of $\mathcal{Y}^{\mu_1\cdots\mu_{k+1}}$, as the ansatz \eqref{eq: general fermion split} would. The transformation is a complicated non-local expression in terms of $\Psi$ which we omit for simplicity, but locally defined in terms of $\mathcal{Y}^{\mu_1\cdots\mu_{k+1}}$ by the transformation rule \eqref{eq:hiddenconformalsymmetry}.

A similar description of hidden conformal symmetry was given for the discrete series scalar fields 
in \cite{Chen:2026kjo,Lindwasser:2026exa}, in terms of a symmetric and traceless conformal Killing tensor. 
There, this description is used to explain the origin of the correlation functions in $\mathrm{dS}_2$ satisfying global conformal Ward identities \cite{Farnsworth:2024yeh}. 
For a treatment of the discrete series fermions in $\mathrm{dS}_2$, where the mass parameter $m_k=-\mathrm{i}(k+1)\sqrt{R/2}$ is imaginary and quantisation requires care, see \cite{Letsios:2025pqo}.

\section{Chiral algebra \label{sec:chiralalgebra}}

In this section, we will describe the general structure of the classical, non-linear, chiral symmetries present in this theory due to holomorphicity of $F(z)$, ignoring global aspects due to boundary conditions. 
This results in an enhancement of local symmetry transformations that do not exist for a fermionic field with a generic mass $m$. 
This description will follow that of the chiral algebra in the discrete series scalar in \cite{Chen:2026kjo}, as well as \cite{Lindwasser:2024qyh,Lindwasser:2025slu}. 
A similar description for the anti-chiral symmetries due to the anti-holomorphicity of $\ol{F}(\zb)$ follows \textit{mutatis mutandis}.

\subsection{Symmetry transformations}

We first note that, because $F(z)$ is holomorphic, any functional $\mathcal{L}(\{F,\pp F,\dots\};z)$ of $F(z)$, all its $z$-derivatives, and arbitrary explicit $z$-dependence, is also holomorphic. 
This basic fact implies an enormous set of symmetries --- one for every functional $\mathcal{L}$. 
For notational simplicity, we will write $F_{n} \equiv \pp^{n}F$ and denote a generic functional with at most $n$ $z$-derivatives (up to integration by parts) by $\mathcal{L}(\{F_n(z)\},z)$.

\paragraph{Massless fermion.} To begin, let us consider first the set of chiral symmetries present in the theory of a massless holomorphic fermion $\psi$ and an anti-holomorphic fermion $\tilde{\psi}$ on flat space, as discussed in \cite{Lindwasser:2024qyh}. 
This theory is conformally invariant and locally equivalent to the discrete series fermion theory with $k=-1$ (it possesses no shift symmetries). 
In this case, the holomorphic current is $\psi(z)$ itself. 
Now, any functional $\mathcal{L}(\{\psi_n(z)\};z)$ (which can be either commuting or anti-commuting) corresponds to an infinitesimal symmetry transformation:
\begin{equation}
    \begin{aligned}
	    &\tilde{\psi}\mapsto \tilde{\psi}+\epsilon\delta_{\mathcal{L}}\tilde{\psi}, &&\delta_{\mathcal{L}}\tilde{\psi}=0,\\
        &\psi\mapsto \psi+\epsilon\delta_{\mathcal{L}}\psi,&& \delta_{\mathcal{L}}\psi=\frac{1}{2}\mathcal{E}(\mathcal{L}),
	\end{aligned}
	\label{eq:masslessfermion sym trans}
\end{equation}
where $\epsilon$ is some infinitesimal parameter which can be commuting or anti-commuting, chosen so that $\epsilon\delta_{\mathcal{L}}\psi$ is anti-commuting. 
The linear operator 
\begin{equation}
	\mathcal{E}(\mathcal{L})\equiv \sum_{i=0}^\infty(-1)^i\pp^i\frac{\pp}{\pp\psi_i(z)}\mathcal{L}\(\{\psi_n(z)\};z\)
\end{equation}
is the Euler operator which gives the Euler--Lagrange equation for $\psi(z)$, as derived from the functional $\mathcal{L}(\{\psi_n(z)\};z)$ interpreted as a $1d$ Lagrangian. 
It vanishes for any value of $\psi(z)$ and its derivatives if and only if $\mathcal{L}$ is a total derivative \cite{Olver:1993}, so that $\mathcal{L}$ and $\mathcal{L}'=\mathcal{L}+\pp f$ generate the same symmetry transformation. 
Thus, the space of symmetries is the space of functionals modulo total derivatives, i.e. the space of $1d$ Lagrangians.

The algebra of symmetry transformations is a graded Lie algebra, which is endowed with a graded commutator 
\begin{equation}
	[A,B]_{-} \equiv AB-(-1)^{|A||B|}BA.
\end{equation}
We take $|A|=0$ if $A$ is commuting and $|A| = 1$ if $A$ anti-commuting. 
The graded commutator between two infinitesimal transformations \eqref{eq:masslessfermion sym trans} associated with Lagrangians $\mathcal{L}$ and $\mathcal{L}'$ is a transformation with respect to a third Lagrangian $\mathcal{L}''$, where
\begin{equation}
    [\delta_{\mathcal{L}},\delta_{\mathcal{L}'}]_{-}=\delta_{\mathcal{L}''},\quad \mathcal{L}''=\frac{1}{4}(-1)^{|\mathcal{L}|}[\mathcal{E}(\mathcal{L}),\mathcal{E}(\mathcal{L}')]_{+}.
    \label{eq:masslessfermioncomm}
\end{equation}
Here, we have defined the graded \textit{anti}-commutator 
\begin{equation}
	[A,B]_{+} \equiv AB+(-1)^{|A||B|}BA.
\end{equation}
We have written $\mathcal{L}''$ in a way to reflect the symmetry of the graded commutator under $(\mathcal{L}\leftrightarrow\mathcal{L}')$, although it can be written more compactly as
\begin{equation}
	\mathcal{L}''= \frac{1}{2}(-1)^{|\mathcal{L}|}\mathcal{E}(\mathcal{L})\mathcal{E}(\mathcal{L}').
\end{equation}

\paragraph{Discrete series fermions.} In appendix \ref{app:invariance of action}, it is shown that for all $k\geq 0$, the action \eqref{eq: dirac action} for the discrete series fermion in $\mathrm{EAdS}_2$ is invariant under the infinitesimal transformation:
\begin{equation}
    \begin{aligned}
	    &\tilde{\psi}\mapsto \tilde{\psi}+\epsilon\delta_{\mathcal{L}}\tilde{\psi}, &&\delta_{\mathcal{L}}\tilde{\psi}=\ii\frac{m_{k}}{4}(-1)^k\Omega^{-1/2}\mathcal{D}_{k}\big(\Omega \,\mathcal{E}(\mathcal{L})\big),\\
        &\psi\mapsto \psi+\epsilon\delta_{\mathcal{L}}\psi,&& \delta_{\mathcal{L}}\psi=\frac{1}{2}(-1)^{k+1}\Omega^{1/2}\mathcal{D}_{k+1}\(\Omega^{-1}\mathcal{E}(\mathcal{L})\),
   	\end{aligned}
\end{equation}
where the Euler operators appearing here are now for functionals of $F$:
\begin{equation}
	\mathcal{E}(\mathcal{L})\equiv \sum_{i=0}^\infty(-1)^i\pp^i\frac{\pp}{\pp F_i(z)}\mathcal{L}(\{F_n(z)\},z).
	\label{eq: higher-order euler}
\end{equation}
These infinitesimal transformations can be written in a more covariant form via \eqref{eq: general fermion split} as an infinitesimal transformation of $\Psi$:
\begin{equation}
    \begin{aligned}
    	\delta_{\mathcal{L}}\Psi=&\DD^{-}_{\mu_1}\cdots\DD^{-}_{\mu_{k+1}}\mathcal{Y}_{\mathcal{L}}^{\mu_1\cdots\mu_{k+1}} \\
    	\mathcal{Y}_{\mathcal{L}}^{z\cdots z}=\frac{1}{2}(-1)^{k+1}
    	&\begin{pmatrix}
    		0\\
    		\Omega^{-1/2}\mathcal{E}(\mathcal{L})
    	\end{pmatrix},\quad \mathcal{Y}_{\mathcal{L}}^{\zb\cdots\zb}=0.
    \end{aligned}
    \label{eq:covsymm}
\end{equation}
This is equivalent to an infinitesimal shift in the component $Y(z)$,
\begin{equation}
    \delta_{\mathcal{L}}Y(z)=\frac{1}{2}(-1)^{k+1}\mathcal{E}(\mathcal{L}),\quad \delta_{\mathcal{L}}\tilde{Y}(\zb)=0.
    \label{eq: Y inf trans}
\end{equation}
These include the linear shift transformations which exist in $\mathrm{dS}_2$ explored in \cite{Bonifacio:2023prb}.
Consider e.g. $\mathcal{L}=p(z) F$.
Then $\mathcal{E}(\mathcal{L})=p(z)$, so by \eqref{eq: Y inf trans} we have $\delta Y\propto p(z)$. 
For $p(z)$ annihilated by $\pp^{2k+2}$, we then have $\delta F =0$ --- these correspond to the $2k+2$-dimensional space of shift symmetries expected in $\mathrm{dS}_2$. 

\subsection{Symmetry algebra}
In the remainder of this section, we will show that for the discrete series fermion theory with $k\geq 0$, the graded commutator $[\delta_{\mathcal{L}},\delta_{\mathcal{L}'}]_{-}$ generates a new transformation $\delta_{\mathcal{L}''}$ associated with the new Lagrangian $\mathcal{L}''$, defined by
\begin{equation}
    [\delta_{\mathcal{L}},\delta_{\mathcal{L}'}]_{-}=\delta_{\mathcal{L}''},\quad \mathcal{L}''=\frac{1}{4}(-1)^{|\mathcal{L}|}[\pp^{k+1}\mathcal{E}(\mathcal{L}),\pp^{k+1}\mathcal{E}(\mathcal{L}')]_{+}.
    \label{eq:Yfermioncomm}
\end{equation}
To show this, we note the similarity between \eqref{eq: Y inf trans} and \eqref{eq:masslessfermion sym trans}, which suggests the following, simple way to determine the graded commutator. 
Consider the transformation \eqref{eq:masslessfermion sym trans} restricted to the subset of functionals $\mathcal{L}(\{\psi_{k+1+n}(z)\};z)$ which depend only on $\psi_{k+1}$, $\psi_{k+2}$, $\dots$, $\psi_{k+1+n}$. 
This is equivalent to considering functionals of $\psi_{k+1}(z)$ rather than $\psi(z)$. 
The Euler operator \eqref{eq: higher-order euler} appearing in \eqref{eq:masslessfermion sym trans} restricted to functionals of $\psi_{k+1}$ simplifies to
\begin{equation}
	\begin{aligned}
	    \mathcal{E}(\mathcal{L})&=\sum_{i=k+1}^\infty(-1)^i\pp^i\frac{\pp}{\pp \psi_i(z)}\mathcal{L}(\{\psi_{k+1+n}(z)\};z) \\
	    &=(-1)^{k+1}\pp^{k+1}\sum_{i=0}^{\infty}(-1)^{i}\pp^i\frac{\pp}{\pp\psi_{k+1+i}(z)}\mathcal{L}(\{\psi_{k+1+n}(z)\};z) \\
	    &=(-1)^{k+1}\pp^{k+1}\mathcal{E}^{k+1}(\mathcal{L}),
    \end{aligned}
\end{equation}
where $\mathcal{E}^{k+1}$ denotes the Euler operator acting on functionals of $\psi_{k+1}(z)$, i.e. where we treat $\psi_{k+1}(z)$ as the function one varies the $1d$ Lagrangian $\mathcal{L}(\{\psi_{k+1+n}(z)\};z)$ with respect to. 
With respect to functionals of $\psi_{k+1}$, $\psi(z)$ transforms as
\begin{equation}
    \delta_{\mathcal{L}}\psi=\frac{1}{2}(-1)^{k+1}\pp^{k+1}\mathcal{E}^{k+1}(\mathcal{L}).
    \label{eq:psirestrictsymtran}
\end{equation}
Now, denoting $Y_{n} \equiv \pp^{n}Y$, we have
\begin{equation}
	\begin{aligned}
	    \delta_{\mathcal{L}}Y_{k+1} &=\frac{1}{2}(-1)^{k+1}\pp^{k+1}\mathcal{E}(\mathcal{L}) \\
	    &=\frac{1}{2}(-1)^{k+1}\pp^{k+1}\sum_{i=0}^\infty(-1)^i\pp^i\frac{\pp}{\pp Y_{2k+2+i}(z)}\mathcal{L}(\{Y_{2k+2+n}(z)\},z),
	\end{aligned}
	\label{eq:Yk+1symtrans}
\end{equation}
which is structurally identical to \eqref{eq:psirestrictsymtran} after relabelling $Y_{k+1}\mapsto\psi$ and $\mathcal{E}\mapsto\mathcal{E}^{k+1}$. 
Because of the equivalence between \eqref{eq:psirestrictsymtran} and \eqref{eq:Yk+1symtrans}, the graded commutator $[\delta_{\mathcal{L}},\delta_{\mathcal{L}'}]_{-}$ for the discrete series fermion with $k\geq 0$ in $\mathrm{EAdS}_2$ is isomorphic to the graded commutator in the massless fermion theory for the subset of functionals $\mathcal{L}(\{\psi_{k+1+n}(z)\},z)$ of $\psi_{k+1}$. 
Thus, the graded Lie algebra of chiral symmetries of the discrete series fermion with $k\geq 0$ is a subalgebra of the graded Lie algebra of chiral symmetries present in the massless fermion theory.
It is a proper subalgebra: For example, $\mathcal L=\psi\partial\psi$ gives $\mathcal E(\mathcal L)=2\partial\psi$, which is not $\partial^{k+1}$ of any functional of $\psi_{k+1},\psi_{k+2},\dots$ for $k\ge0$.

We now use this isomorphism to compute the graded commutator $[\delta_{\mathcal{L}},\delta_{\mathcal{L}'}]_{-}$ for the transformations \eqref{eq: Y inf trans}, by taking \eqref{eq:masslessfermioncomm} and restricting $\mathcal{L}$ and $\mathcal{L}'$ to be functionals of $\psi_{k+1}(z)$, and making the replacements $\psi_{k+1}(z)\mapsto F(z)$ and $\mathcal{E}^{k+1}\mapsto \mathcal{E}$, we arrive at \eqref{eq:Yfermioncomm}.
Here we have used the fact that $F(z)=Y_{2k+2}(z)$, see \eqref{eq: field strength parameterisation}.

\section{Concluding remarks}

In this work, we extended the analysis of \cite{Chen:2026kjo, Lindwasser:2026exa} to the fermionic theories in AdS$_2$ studied in \cite{Bonifacio:2023prb, Letsios:2025pqo}.
To make the parallel with $2d$ CFTs explicit, we expressed the general solution \eqref{eq: holomorphic splitting} to the classical equation of motion in terms of two arbitrary (anti-)holomorphic functions.
This allowed us to quantise the theory using simple Laurent series mode expansions, enabling straightforward computations of field strength correlation functions.
These correlation functions are reminiscent of those of conformal primaries in flat space CFTs with boundary.
By identifying the arbitrary functions with components of symmetric and traceless conformal Killing spinor--tensors, we trace this connection to a hidden conformal symmetry of the equation of motion which is made manifest at the level of these tensors.
Finally, we studied the associated chiral algebras and showed that they are proper subalgebras of the chiral algebra of the ordinary Majorana fermion on two-dimensional Minkowski space.

\hspace{2em} Some future directions are noteworthy. 
In two dimensions bosons and fermions are connected by local bosonisation maps.
For instance, the free massive Dirac fermion is equivalent to the sine-Gordon model at a special coupling \cite{PhysRevD.11.2088,PhysRevD.36.2408}.
The sine-Gordon model is famously integrable, with an infinite sequence of conserved currents \cite{Yoon:1975ma}. 
At the fermionic mass values $m=m_k$ in $\mathrm{AdS}_2$, the corresponding sine-Gordon model will then have, in addition, (anti-)holomorphic currents.
It will be interesting to consider how this manifests from the bosonic perspective.

The $k=0,1$ holomorphic scalars and fermions organise into supersymmetric multiplets in the Green--Schwarz superstring worldsheet theory as coordinate fluctuations around various string solutions in $\mathrm{AdS}_5\times S^5$ \cite{Metsaev:1998it,Drukker:2000ep,Giombi:2017cqn, Beccaria:2026ffm}. The holomorphicity present in the quadratic fluctuations is suggestive, and further investigation as to what extent it is relevant to the connection with $\mathcal{N}=4$ super Yang--Mills theory is needed.
Beyond the free theory, the Abelian $k=1$ shift can be deformed into a
\textit{non-Abelian} algebra: For a brane, the broken isometries close onto the unbroken
ones \cite{Goon:2012dy, Bonifacio:2021mrf, Garoffolo:2025igz}, e.g.\
$\mathfrak{so}(2,2) \to \mathfrak{so}(2,1)$ for AdS$_2 \subset$ AdS$_3$.
These are field-dependent, non-linear extensions of the shift symmetries in the free theories we have described here, which were described in \cite{Bonifacio:2018zex, Bonifacio:2023prb}. 

The hidden conformal symmetry of the discrete series scalars and fermions in (A)dS$_{2}$ is closely related to the fact that they correspond to so-called singleton representations \cite{Siegel:1988gd, Angelopoulos:1997ij, Laoues:1998ik, Angelopoulos:1999bz}.
Action formulations for these broadly fall into two categories: Those which make their field content explicit while breaking manifest Lorentz covariance, and those which make Lorentz covariance manifest while keeping their field content hidden.
The massive scalar and fermion actions with specific mass values are an example of the latter.
From this, it is clear that alternative formulations are preferred to make chirality front-and-centre. 
For scalar singletons in two dimensions, such actions were derived in \cite{Chen:2025xlo}.
In \cite{hifj}, the suggestion that these describe the degrees of freedom of the discrete series theories is taken seriously, and their hidden conformal symmetry is explained in terms of ordinary conformal symmetry acting on the higher-order chiral scalar theories.
This analysis should extend to the fermionic theories discussed in this manuscript.
In particular, the higher-order chiral fermions should correspond to the boundary degrees of freedom of fermionic versions of higher-spin Chern--Simons theory in AdS$_{3}$ and be described by actions with even-order Bol operators.
It would be interesting to study these.

\section*{Acknowledgements}

It is our pleasure to thank Kurt Hinterbichler, Massimo Porrati, Samanta Saha, and Thomas Yan for illuminating discussions.
CYRC would further like to thank Euihun Joung and Karapet Mkrtchyan for collaboration on related work.

The research of CYRC was funded by Grant 115L104044 (Max Planck--IAS--NTU Center) supported by National Taiwan University under the framework of Higher Education SPROUT Project by the Ministry of Education in Taiwan. 
The research of LWL is supported by the Taiwan NSTC Grant No. 113-2811-M-002-167-MY3 and the Yushan Young Fellowship.

We acknowledge the use of ChatGPT-6 (OpenAI) and Opus 5.5 (Anthropic) to assist with editing. 
The authors take full responsibility for the content of the paper.

\bibliography{Bibliography}

\appendix

\section{Fermion conventions \label{app: fermion conventions}}

In this appendix, we fix the conventions for how we describe fermions in $\text{AdS}_{2}$ in both global and complex coordinates. 

\subsection{Strip: Global coordinates \label{sapp: global conventions}}

Let us introduce the zweibein basis 
\begin{equation}
    e^{\ul{0}} = \frac{L}{\cos(\rho)} \dd \tau,\quad e^{\ul{1}} = \frac{L}{\cos(\rho)} \dd \rho,
\end{equation}
for which the only non-zero component of the torsion-less connection one-form is
\begin{equation}
    \omega^{\ul{0}}{}_{\ul{1}} = \tan(\rho) \dd \tau.
\end{equation}
This spacetime has three Killing vectors,
\begin{subequations}
\begin{align}
    &K_0=\pp_\tau,\\
    &K_1=\cos(\rho)\sin(\tau)\pp_\rho+\sin(\rho)\cos(\tau)\pp_\tau,\\
    &K_2=\cos(\rho)\cos(\tau)\pp_\rho-\sin(\rho)\sin(\tau)\pp_\tau.
\end{align}
\end{subequations}
We take $\gamma$-matrices to satisfy
\begin{equation}
    \{\gamma_{\mu},\gamma_{\nu}\} = 2g_{\mu\nu} \quad \implies \quad \{\gamma_{\ul{a}},\gamma_{\ul{b}}\} = 2g_{\ul{ab}}.
\end{equation}
Following \cite{Freedman:2012zz}, we pick
\begin{equation}
    \gamma^{\ul{0}} = \ii\sigma_{2} = 
    \begin{pmatrix}
        0 & 1 \\
        -1 & 0
    \end{pmatrix}, \quad 
    \gamma^{\ul{1}} = \sigma_{1} = 
    \begin{pmatrix}
        0 & 1 \\
        1 & 0 
    \end{pmatrix}.
\end{equation}
In two dimensions, there is only one other element in the Clifford algebra, whose only non-trivial component is
\begin{equation}
    \gamma^{\ul{01}} = \sigma_{3} = 
    \begin{pmatrix}
        1 & 0 \\
        0 & -1 
    \end{pmatrix}.
\end{equation}
This is related to the highest-weight element
\begin{equation}
    \gamma_{*} = -\gamma_{\ul{0}}\gamma_{\ul{1}} = \sigma_{3},
\end{equation}
with which one can define the projection operators
\begin{equation}
    \PP_{\pm} = \frac{1 \pm \gamma_{*}}{2}.
\end{equation}
Finally, we pick our charge conjugation matrix to be
\begin{equation}
    C = \sigma_{2},
\end{equation}
which corresponds to $C_{-}$ in \cite{Freedman:2012zz} with $t_{0} = 1$ and $t_{1} = -1$.

The full covariant derivative acts on a spinor--tensor as
\begin{equation}
    \DD_{\mu}  = \nabla_{\mu}   + \frac{1}{4}\omega^{\ul{ab}}{}_{\mu} \gamma_{\ul{ab}} ,
    \label{eq: spinor derivative def}
\end{equation}
where $\nabla$ is the ordinary covariant derivative acting on a tensor
\begin{equation}
    \nabla_{\mu} V^{\ul{a}} = \pp_{\mu}V^{\ul{a}}  + \omega^{\ul{a}}{}_{\ul{b}\mu}V^{\ul{b}}.
\end{equation}
The components of this are
\begin{subequations}
    \begin{align}
        \DD_{0} &= \nabla_{0}  - \frac{1}{2} \tan(\rho)\sigma_{3} , \\
        \DD_{1}  &= \nabla_{1} .
    \end{align}
\end{subequations}
When acting on a rank-$s$ spinor--tensor $\Psi_{\lambda_{1} \dots \lambda_{s}}$, we also have
\begin{equation}
    [\DD_{\mu},\DD_{\nu}] \Psi_{\lambda_{1} \dots \lambda_{s}} = - \frac{1}{2L^{2}} \gamma_{\mu\nu} \Psi_{\lambda_{1} \dots \lambda_{s}} - \frac{2}{L^{2}} \sum_{i=1}^{s} g_{\lambda_{i}[\mu|}\Psi_{\lambda_{1}\dots \lambda_{i-1}|\nu]\lambda_{i+1}\dots \lambda_{s}}.
    \label{eq: cov deriv commutation}
\end{equation}

\subsection{Upper-half plane: Complex coordinates \label{sapp: UHP conventions}}

We introduce the zweibein basis
\begin{equation}
    e^{\ul{z}} = \Omega \,\dd z,\quad e^{\ul{\zb}} = \Omega\,\dd\zb,
\end{equation}
for which the only non-zero component of the torsion-less connection one-form is
\begin{equation}
    \omega^{\ul{z\zb}} = \frac{2\pp \Omega}{\Omega} \dd z- \frac{2 \ol{\pp}\Omega}{\Omega} \dd \zb.
\end{equation}
In these coordinates the spacetime has the three Killing vectors,
\begin{subequations}
\label{eq:AdSisometries}
    \begin{align}
        \ell_{+1}&=-z^2\pp-\zb^2\ol\pp,\\
        \ell_0&=-z\pp-\zb\ol\pp,\\
        \ell_{-1}&=-\pp-\ol\pp.
    \end{align}
\end{subequations}
We choose the $\gamma$-matrices to be 
\begin{equation}
\label{eq:Euclideangammas}
    \gamma_{\ul{z}} = 
    \begin{pmatrix}
        0 & 1 \\
        0 & 0
    \end{pmatrix},\quad 
    \gamma_{\ul{\zb}} = 
    \begin{pmatrix}
        0 & 0 \\
        1 & 0
    \end{pmatrix},
\end{equation}
so that the remaining non-trivial element of the Clifford algebra is
\begin{equation}
    \gamma_{\ul{z\zb}} = \frac{1}{2}\sigma_{3} = \frac{1}{2}
    \begin{pmatrix}
        1 & 0 \\
        0 & -1
    \end{pmatrix}.
\end{equation}

In components, the spinor covariant derivative \eqref{eq: spinor derivative def} is
\begin{subequations}
    \begin{align}
        \DD \equiv \DD_{z}   &= \(\nabla_{z} + \frac{\pp \Omega}{2 \Omega} \sigma_{3} \) \\
        \ol{\DD} \equiv \DD_{\zb}  &= \(\nabla_{\zb} - \frac{\ol{\pp}\Omega}{2\Omega} \sigma_{3}\) ,
    \end{align}
\end{subequations}
with
\begin{equation}
    \nabla_{z}  = \(\pp - h \frac{2\pp\Omega}{\Omega}\) ,
\end{equation}
when acting on a rank-$h$ holomorphic tensor. 
It will turn out to be convenient to define the projectors to be
\begin{equation}
    \PP \equiv \PP_{z} = \PP_{-},\quad \ol{\PP} \equiv \PP_{\zb} = \PP_{+},
\end{equation}
which satisfy
\begin{equation}
    \begin{aligned}
        &\PP = \gamma_{\ul\zb}\gamma_{\ul z},\quad \ol{\PP} = \gamma_{\ul z} \gamma_{\ul \zb} \\
        &\PP \gamma_{\ul \zb} = \gamma_{\ul \zb},\quad \ol{\PP} \gamma_{\ul z} = \gamma_{\ul z}.
    \end{aligned}
    \label{eq: gamma projections}
\end{equation}

\section{Holomorphicity of field strength \label{app: field strength holomorphicity}}

Let us see how the conservation of the field strength components \eqref{eq:complexconservation} works explicitly.

First note that since the covariant derivative commutes with the projectors, conservation is trivial for the $\ol{\PP}$-projection as this vanishes on-shell.
We therefore just need to focus on the $\PP$-projection.
To that end, let us first note that the commutation relations \eqref{eq: cov deriv commutation} applied to the case at hand reads
\begin{equation}
    \[\ol{\DD},\DD\]\DD^{\alpha} \Psi = - \frac{\Omega^{2}}{4L^{2}}\(2\alpha \mathrm{I} - \sigma_{3}\) \DD^{\alpha}\Psi,
\end{equation}
so that
\begin{equation}
    \begin{aligned}
        \[\ol{\DD},\DD^{\alpha+1}\] \Psi &= \sum_{n=0}^{\alpha} \DD^{\alpha-n}\[\ol{\DD},\DD\]\DD^{n} \Psi \\
        &=   - \frac{\Omega^{2}}{4L^{2}} \sum_{n=0}^{\alpha} \(2n \, \text{I} - \sigma_{3}\)\DD^{\alpha} \Psi \\
        &=   - \frac{\Omega^{2}}{4L^{2}}(\alpha+1)\(\alpha\, \text{I} - \sigma_{3}\)\DD^{\alpha} \Psi.
    \end{aligned}
\end{equation}

By projecting with $\PP$ and making repeated use of the Dirac equation \eqref{eq: dirac eq} with $\gamma_{z}$ and $\gamma_{\zb}$ acting on the left (which appeared in \eqref{eq: field strength on-shell projections}), we then find that for the discrete series fermion  
\begin{align}
    \PP\ol{\DD}\mathcal{F}_{z \dots z} &= \PP \ol{\DD} \DD^{k+1}\Psi \nonumber \\
    &= \DD^{k+1}\ol{\DD}\(\PP\Psi\) - \frac{\Omega^{2}}{4L^{2}} (k+1)\(k \, \mathrm{I} - \sigma_{3}\)\DD^{k}\(\PP \Psi\) \nonumber \\
    &= - \ii \frac{m_{k}}{2} \DD^{k+1}\( \sigma_{3}\gamma_{\zb}\Psi\) - \frac{\Omega^{2}}{4L^{2}} (k+1)\(k \, \mathrm{I} - \sigma_{3}\)\DD^{k}\(\PP \Psi\) \nonumber \\
    &= - \ii \frac{m_{k}}{2} \sigma_{3} \DD^{k}\[\gamma_{\zb}\DD\( \ol{\PP}\Psi\)\] - \frac{\Omega^{2}}{4L^{2}} (k+1)\(k \, \mathrm{I} - \sigma_{3}\)\DD^{k}\(\PP \Psi\) \\
    &= \(- \ii \frac{m_{k}}{2}\)^{2} \sigma_{3} \DD^{k}\(\gamma_{\zb}\sigma_{3}\gamma_{z}\Psi\) - \frac{\Omega^{2}}{4L^{2}} (k+1)\(k \, \mathrm{I} - \sigma_{3}\)\DD^{k}\(\PP \Psi\) \nonumber \\
    &= \frac{\Omega^{2}}{4L^{2}}\[m_{k}^{2}L^{2} - (k+1)\(k \, \mathrm{I} - \sigma_{3}\)\] \DD^{k}\(\PP\Psi\), \nonumber 
\end{align}
which vanishes on-shell, as required. 

\section{Fermion holomorphic splitting \label{app: holomorphic splitting}}

Our central result is that the phase space of the shift-symmetric fermion in AdS$_{2}$ is fully parameterised by the holomorphic splitting \eqref{eq: holomorphic splitting}.
We will show this directly and also present a derivation based on the covariant tensor introduced in section~\ref{sec:Hiddenconformalsymmetry}. 

\subsection{Non-covariant}

To show this directly, we will focus on just the equation for the anti-holomorphic part $\tilde{\psi}$.

First note that 
\begin{equation}
    \begin{aligned}
        \(\pp - \frac{1}{2(z-\zb)}\)\tilde{\psi}_{Y} &= \(\pp + \frac{\pp\Omega}{2\Omega}\)\tilde{\psi}_{Y} \\
        &= \Omega^{-1/2}\pp\(\Omega^{1/2}\tilde{\psi}_{Y}\) \\
        &= \Omega^{-1/2}\pp\bigg[\Omega\ol{\mathcal{D}}_{k+1}\big(\Omega^{-1}\tilde{Y}(\zb)\big) \bigg] - \ii\frac{m_{k}}{2}  \Omega^{-1/2}\pp\mathcal{D}_{k}\big(\Omega \,Y(z)\big).
    \end{aligned}
\end{equation}
Let us consider these terms separately.
First, we have
\begin{equation}
    \begin{aligned}
        \pp\[\Omega\ol{\mathcal{D}}_{k+1}\big(\Omega^{-1}\tilde{Y}(\zb)\big) \] &= \pp \[\Omega^{-k-1}\ol{\pp}^{k+1}\big(\Omega^{k+1}\tilde{Y}(\zb)\big)\] \\
        &= (k+1)\(\frac{\ii}{2L}\) \Omega^{-k}\[\Omega^{-1}\ol{\pp}^{k+1}\big(\Omega^{k+2}\tilde{Y}(\zb)\big)-\ol{\pp}^{k+1}\big(\Omega^{k+1}\tilde{Y}(\zb)\big) \] \\
        &= (k+1)\(\frac{\ii}{2L}\) \Omega^{-k}\bigg[\Omega^{-1}\ol{\pp}^{k+1}\big(\Omega^{k+2}\tilde{Y}(\zb)\big)\\
        &\quad - \sum_{n=k}^{k+1}(-)^{k-n+1} \frac{(k+1)!}{(k+1-n)!n!(n-k)!}\ol{\pp}^{n}\big(\Omega^{k+2}\tilde{Y}(\zb)\big) \(\frac{-\ii}{2L}\)^{k+1-n}\Omega^{k-n}\bigg] \\
        &=  \(\frac{k+1}{2L}\)^{2} \Omega^{-k} \ol{\pp}^{k}\big(\Omega^{k+2}\tilde{Y}(\zb)\big) \\
        &= \(\frac{m_{k}}{2}\)^{2} \Omega \ol{\mathcal{D}}_{k}\big(\Omega \tilde{Y}(\zb)\big).
    \end{aligned}
\end{equation}
Next, note that
\begin{align}
    \pp\mathcal{D}_{k}\big(\Omega Y(z)\big) &= \pp\[\Omega^{-k-1}\pp^{k}\big(\Omega^{k+2}Y(z)\big)\] \nonumber \\
    &= \sum_{n=0}^{k} \binom{k}{n} (k-n)!\(\frac{\ii}{2L}\)^{k-n}\pp\[\Omega^{-n}\pp^{n}\big(\Omega^{k+1}Y(z)\big)\] \\
    &= \sum_{n=0}^{k} \frac{k!}{n!} \(\frac{\ii}{2L}\)^{k-n}\[\Omega^{-n}\pp^{n+1}\big(\Omega^{k+1}Y(z)\big) - n \(\frac{\ii}{2L}\) \Omega^{-n+1}\pp^{n}\big(\Omega^{k+1}Y(z)\big)\] \nonumber \\
    &= \Omega^{-k}\pp^{k+1}\big(\Omega^{k+1} Y(z) \big) \nonumber \\
    &= \Omega^{2}\mathcal{D}_{k+1}\big(\Omega^{-1}Y(z) \big). \nonumber 
\end{align}
Putting everything together, 
\begin{align}
    \(\pp-\frac{1}{2(z-\zb)}\)\tilde{\psi}_{Y} &= \(\frac{m_{k}}{2}\)^{2} \Omega^{1/2}\ol{\mathcal{D}}_{k}\big(\Omega \tilde{Y}(\zb)\big)  - \ii\frac{m_{k}}{2}  \Omega^{3/2}\mathcal{D}_{k+1}\big(\Omega^{-1}Y(z) \big) \nonumber \\
    &=  -\ii \frac{m_{k}}{2}  \Omega \[\Omega^{1/2}\mathcal{D}_{k+1}\big(\Omega^{-1}Y(z) \big) +  \ii \frac{m_{k}}{2} \Omega^{-1/2}\ol{\mathcal{D}}_{k}\big(\Omega \tilde{Y}(\zb)\big)\] \nonumber \\
    &= -\ii \frac{m_{k}}{2} \Omega \psi_{Y} \\
    &=  \frac{m_{k}L}{z-\zb} \psi_{Y}, \nonumber 
\end{align}
which is precisely the equation of motion, as required.
The equation of motion for the holomorphic sector follows similarly (or by complex conjugation).

\subsection{Covariant}

We will now prove that the Euclidean signature massive Dirac equation \eqref{eq: dirac eq} at the discrete set of masses \eqref{eq: general discrete series masses},
\begin{equation}
	\(\gamma^\mu \mathrm{D}_\mu-(k+1)\sqrt{\frac{R}{2}}\gamma_*\)\Psi = 0
\end{equation}
is solved by \eqref{eq: general fermion split} for any traceless and symmetric spinor--tensor $\mathcal{Y}^{\mu_1\cdots\mu_{k+1}}$ satisfying the additional conditions \eqref{eq: YCKST}.
That this is a solution to the massive Dirac equation follows from the following identity:

\begin{identity}
On a two-dimensional, Euclidean manifold with constant Ricci scalar $R$, any symmetric and traceless rank $k+1$ spinor--tensor $\mathcal{Y}^{\mu_1\cdots\mu_{k+1}}$ satisfies the intertwining relation
\begin{equation}
	\begin{aligned}
		&\(\gamma^\nu \mathrm{D}_\nu-(k+1)\sqrt{\frac{R}{2}}\gamma_*\) \DD^{-}_{\mu_1}\cdots \DD^{-}_{\mu_{k+1}}\mathcal{Y}^{\mu_1\cdots\mu_{k+1}}= \\
		&\DD^{-}_{\mu_1}\cdots\DD^{-}_{\mu_{k+1}}\gamma_{\mu_{k+2}}\DD^{(\mu_1}\mathcal{Y}^{\mu_2\cdots\mu_{k+2})_T} +\frac{1}{2}\(\gamma^\nu \mathrm{D}_\nu-(k+1)\sqrt{\frac{R}{2}}\gamma_*\)\DD^{-}_{\mu_1}\cdots\DD^{-}_{\mu_{k+1}}\gamma_{\mu_{k+2}}\gamma^{(\mu_1}\mathcal{Y}^{\mu_2\cdots\mu_{k+2})_T}.
	\end{aligned}
	\label{eq: Y main identity}
\end{equation}
\end{identity}
\begin{proof}
The proof of this identity follows in three steps. 

First, we note that in two dimensions, it is possible to decompose $\mathcal{Y}^{\mu_1\cdots\mu_{k+1}}$ into two parts,
\begin{equation}
	\mathcal{Y}^{\mu_1\cdots\mu_{k+1}}=\frac{1}{2}\gamma^{(\mu_1}\mathcal{B}^{\mu_2\cdots\mu_{k+1})}+\frac{1}{2}\gamma_{\mu_{k+2}}\gamma^{(\mu_1}\mathcal{Y}^{\mu_2\cdots\mu_{k+2})_T},\quad \mathcal{B}^{\mu_1\cdots\mu_k}\equiv\gamma_{\nu}\mathcal{Y}^{\nu\mu_1\cdots\mu_{k}}
	\label{eq:Ydecomp}
\end{equation}
This follows directly from expanding
\begin{equation}
	\gamma^{(\mu_1}\mathcal{Y}^{\mu_2\cdots\mu_{k+2})_T}\equiv\gamma^{(\mu_1}\mathcal{Y}^{\mu_2\cdots\mu_{k+2})}-\frac{1}{2}g^{(\mu_1\mu_2}\gamma_\nu\mathcal{Y}^{\nu\mu_3\cdots\mu_{k+2})}.
\end{equation}
Note that since $\mathcal{Y}^{\mu_1\cdots\mu_{k+1}}$ is symmetric and traceless, $\mathcal{B}^{\mu_1\cdots\mu_k}$ is symmetric and $\gamma$-traceless. 
The second term in the decomposition \eqref{eq:Ydecomp} is zero if $\mathcal{Y}^{\mu_1\cdots\mu_{k+1}}$ satisfies \eqref{eq: YCKST}, so we will focus on the $\mathcal{B}^{\mu_1\cdots\mu_k}$ term in \eqref{eq:Ydecomp} for now. 

The next step is to show that 
\begin{equation}
	\label{eq:Bidentity}
	\DD^{-}_{\mu_1}\cdots\DD^{-}_{\mu_{k+1}}\frac{1}{2}\gamma^{(\mu_1}\mathcal{B}^{\mu_2\cdots\mu_{k+1})}=\frac{1}{2}\(\gamma^\mu \mathrm{D}_\mu-(k+1)\sqrt{\frac{R}{2}}\gamma_*\)\DD_{\mu_1}\cdots \DD_{\mu_k}\mathcal{B}^{\mu_1\cdots\mu_k}.
\end{equation}
This follows recursively by expanding $\DD^{-}$ given in \eqref{eq: general fermion split} on the left-hand side, using the fact that $\mathcal{B}^{\mu_1\cdots\mu_k}$ is $\gamma$-traceless. 
For instance, 
\begin{equation}
	\DD^{-}_{\mu_{k+1}}\gamma^{(\mu_1}\mathcal{B}^{\mu_2\cdots\mu_{k+1})}=\gamma^{(\mu_1}\DD_{\mu_{k+1}}\mathcal{B}^{\mu_2\cdots\mu_{k+1})}-2\sqrt{\frac{R}{8}}\gamma_*\mathcal{B}^{\mu_1\cdots\mu_k}.
\end{equation}
Iterating this $k$ more times gives \eqref{eq:Bidentity}. Acting on \eqref{eq:Bidentity} with the massive Dirac operator therefore gives
\begin{equation}
	\(\gamma^\nu \mathrm{D}_\nu-(k+1)\sqrt{\frac{R}{2}}\gamma_*\)\DD^{-}_{\mu_1}\cdots\DD^{-}_{\mu_{k+1}}\frac{1}{2}\gamma^{(\mu_1}\mathcal{B}^{\mu_2\cdots\mu_{k+1})}\nonumber=\frac{1}{2}\((\gamma^\nu\DD_\nu)^2+\frac{(k+1)^2}{2}R\)\DD_{\mu_1}\cdots \DD_{\mu_k}\mathcal{B}^{\mu_1\cdots\mu_k}.
\end{equation}
Now, the remaining step is to show the following identity,
\begin{equation}
	\label{eq:BtoCKST}
	\((\gamma^\nu\DD_\nu)^2+\frac{(k+1)^2}{2}R\)\DD_{\mu_1}\cdots \DD_{\mu_k}\mathcal{B}^{\mu_1\cdots\mu_k}=2\DD^{-}_{\mu_1}\cdots\DD^{-}_{\mu_{k+1}}\gamma_{\mu_{k+2}}\DD^{(\mu_1}\mathcal{Y}^{\mu_2\cdots\mu_{k+2})_T}.
\end{equation}
Combining this with \eqref{eq:Ydecomp} and \eqref{eq:Bidentity}, yields \eqref{eq: Y main identity}. 

Let us prove \eqref{eq:BtoCKST}. 
First, note that $(\gamma^\nu\DD_\nu)^2=\DD^\nu\DD_\nu-R/4$. 
We start by commuting one of these covariant derivatives through the $k$-fold divergences:
\begin{equation}
	\DD^\nu\DD_\nu\DD_{\mu_1}\cdots\DD_{\mu_k}\mathcal{B}^{\mu_1\cdots\mu_k}=\DD^\nu \DD_{\mu_1}\cdots\DD_{\mu_k}\DD_\nu\mathcal{B}^{\mu_1\cdots\mu_k}-\frac{k(k+2)}{4}R\DD_{\mu_1}\cdots\DD_{\mu_k}\mathcal{B}^{\mu_1\cdots\mu_k}.
\end{equation}
Using this, the left-hand side of \eqref{eq:BtoCKST} simplifies to
\begin{equation}
	\((\gamma^\nu\DD_\nu)^2+\frac{(k+1)^2}{2}R\)\DD_{\mu_1}\cdots \DD_{\mu_k}\mathcal{B}^{\mu_1\cdots\mu_k}=\DD^\nu \DD_{\mu_1}\cdots\DD_{\mu_k}\DD_\nu\mathcal{B}^{\mu_1\cdots\mu_k}+\frac{(k+1)^2}{4}R\DD_{\mu_1}\cdots\DD_{\mu_k}\mathcal{B}^{\mu_1\cdots\mu_k}.
\end{equation}
The first term on the right-hand side requires some additional care. 
Explicitly using complex coordinates $(z,\zb)$, this term expands into four parts
\begin{equation}
	\DD^\nu \DD_{\mu_1}\cdots\DD_{\mu_k}\DD_\nu\mathcal{B}^{\mu_1\cdots\mu_k}= \DD_z^{k+1}\gamma_z\DD^z\mathcal{Y}^{z\cdots z}+\DD_{\zb}^{k+1}\gamma_{\zb}\DD^{\zb}\mathcal{Y}^{\zb\cdots \zb} +\DD_{\zb}\DD_z^k\gamma_z\DD^{\zb}\mathcal{Y}^{z\cdots z}+\DD_z\DD_{\zb}^k\gamma_{\zb}\DD^{z}\mathcal{Y}^{\zb\cdots\zb}
\end{equation}
We recognise the first two terms on the right-hand side as
\begin{equation}
	\DD_{\mu_1}\cdots\DD_{\mu_{k+1}}\gamma_{\mu_{k+2}}\DD^{(\mu_1}\mathcal{Y}^{\mu_2\cdots\mu_{k+2})_T}= \DD_z^{k+1}\gamma_z\DD^z\mathcal{Y}^{z\cdots z}+\DD_{\zb}^{k+1}\gamma_{\zb}\DD^{\zb}\mathcal{Y}^{\zb\cdots \zb} 
\end{equation}
in explicit coordinates. 
Furthermore, because $\gamma_{\mu_{k+2}}\DD^{(\mu_1}\mathcal{Y}^{\mu_2\cdots\mu_{k+2})_T}$ is $\gamma$-traceless, we can replace all covariant divergences $\DD_{\mu_i}\mapsto\DD^{-}_{\mu_i}$. 
Using the fact that $\gamma_z^2=0=\gamma_{\zb}^2$, the remaining terms can then be rewritten as
\begin{equation}
	\begin{aligned}
		g^{\zb z}\DD_{\zb}\DD_z\gamma_z\DD_z^k\mathcal{Y}^{z\cdots z}+g^{z\zb}\DD_z\DD_{\zb}\gamma_{\zb}\DD_{\zb}^k\mathcal{Y}^{\zb\cdots\zb}&=\frac{1}{2}(\gamma^\nu\DD_\nu)^2(\gamma_z\DD_z^k\mathcal{Y}^{z\cdots z}+\gamma_{\zb}\DD_{\zb}^k\mathcal{Y}^{\zb\cdots\zb}) \\
		&=\frac{1}{2}(\gamma^\nu\DD_\nu)^2\DD_{\mu_1}\cdots\DD_{\mu_k}\mathcal{B}^{\mu_1\cdots\mu_k}
	\end{aligned}
\end{equation}
We therefore have
\begin{equation}
	\label{eq:Bidentity2}
	\DD^\nu \DD_{\mu_1}\cdots\DD_{\mu_k}\DD_\nu\mathcal{B}^{\mu_1\cdots\mu_k}=\DD^{-}_{\mu_1}\cdots\DD^{-}_{\mu_{k+1}}\gamma_{\mu_{k+2}}\DD^{(\mu_1}\mathcal{Y}^{\mu_2\cdots\mu_{k+2})_T}+\frac{1}{2}(\gamma^\nu\DD_\nu)^2\DD_{\mu_1}\cdots\DD_{\mu_k}\mathcal{B}^{\mu_1\cdots\mu_k}
\end{equation}
Using \eqref{eq:Bidentity2}, we get \eqref{eq:BtoCKST}.
The result \eqref{eq: Y main identity} follows.
\end{proof}

\section{Field strength holomorphic splitting \label{app: field strenght splitting}}

Let us see how \eqref{eq: field strength parameterisation} materialises.
More specifically, the claim is that the holomorphic splitting \eqref{eq: holomorphic splitting} is also associated with a particularly simple form for the field strength:
\begin{equation}
	\mathcal{F}_{z\cdots z}\big|_{Y} = \(\DD^{+}\)^{k+1}\Psi_{Y} = 
	\begin{pmatrix}
		0 \\
		\Omega^{-1/2}\pp^{2(k+1)} Y(z)
	\end{pmatrix}.
	\label{eq: field strength holomorphic splitting}
\end{equation}

To see this, let us first define the shifted covariant derivative operator 
\begin{equation}
	\nabla_{(\mu)} = \nabla + \mu \frac{\pp\Omega}{\Omega},
\end{equation}
such that
\begin{equation}
	\nabla_{(\mu)}^{\alpha}\cdot = \(\pp + [\mu-2(\alpha-1)]\frac{\pp\Omega}{\Omega}\) \dots \(\pp + \mu\frac{\pp\Omega}{\Omega}\).
\end{equation}
Then, note that
\begin{subequations}
	\begin{align}
		\mathcal{D}_{\alpha} &= \Omega^{-(\alpha+1)} \pp^{\alpha} \[\Omega^{\alpha+1} \cdot\] \\
		\nabla_{(\mu)}^{\alpha} &= \Omega^{\alpha-1-\mu}\pp^{\alpha}\[\Omega^{\mu+1-\alpha}\cdot\].
	\end{align}
\end{subequations}
We will also make liberal use of
\begin{equation}
	\pp^{n}\Omega^{\alpha} = \(\frac{\ii}{2L}\)^{n}\Omega^{\alpha+n}
	\begin{cases}
		\frac{(\alpha+n-1)!}{(\alpha-1)!}, & \alpha >0 \\
		(-)^{n} \frac{(-\alpha)!}{(-\alpha-n)!} , & \alpha \leq 0.
	\end{cases}
\end{equation}
We can now write the spinor covariant derivative as
\begin{equation}
	\DD^{\alpha} = \nabla_{(-1/2)}^{\alpha}\PP +\nabla_{(+1/2)}^{\alpha}\ol{\PP}.
\end{equation}
Let us now consider the $\ol{\PP}$ projection:
\begin{equation}
	\ol{\PP}\mathcal{F}_{z\cdots z}\big|_{Y} = \nabla_{(+1/2)}^{k+1} \ol{\PP}\Psi_{Y}+ \ii\frac{m_{k}}{2} \sigma_{3} \nabla_{(-1/2)}^{k} \gamma_{z}\Psi_{Y} .
\end{equation}
With a slight abuse of notation, the non-trivial component of this is 
\begin{equation}
	\begin{aligned}
		\ol{\PP}\mathcal{F}_{z\cdots z}\big|_{Y} &\ni \nabla_{(+1/2)}^{k+1}\tilde{\psi}_{Y} +  \ii\frac{m_{k}}{2} \Omega\nabla_{(-1/2)}^{k} \psi_{Y} \\
		&= - \ii\frac{m_{k}}{2} \(\hat{\mathrm{O}}_{1} - \hat{\mathrm{O}}_{2}\)Y(z)  + \[ \hat{\mathrm{O}}_{3}  - \(\frac{m_{k}}{2}\)^{2} \hat{\mathrm{O}}_{4} \]\tilde{Y}(\zb), 
	\end{aligned}
\end{equation}
where
\begin{subequations}
	\begin{align}
		\hat{\mathrm{O}}_{1} \cdot &= \nabla_{(+1/2)}^{k+1} \[\Omega^{-1/2} \mathcal{D}_{k}\(\Omega \cdot \)\] \\
		\hat{\mathrm{O}}_{2} \cdot &= \Omega \nabla_{(-1/2)}^{k} \[\Omega^{1/2} \mathcal{D}_{k+1}\(\Omega^{-1}\cdot \)\] \\
		\hat{\mathrm{O}}_{3} \cdot &= \nabla_{(+1/2)}^{k+1} \[\Omega^{1/2}\ol{\mathcal{D}}_{k+1}\(\Omega^{-1}\cdot \)\] \\
		\hat{\mathrm{O}}_{4} \cdot &= \Omega \nabla_{(-1/2)}^{k}\[\Omega^{-1/2} \ol{\mathcal{D}}_{k} \(\Omega \cdot \)\].
	\end{align}
\end{subequations}
Let's consider each term separately.
For the operators acting on $\tilde{Y}(\zb)$, we have
\begin{equation}
	\begin{aligned}
		\hat{\mathrm{O}}_{3}\tilde{Y}(\zb)& = \Omega^{k-1/2}\pp^{k+1}\[\Omega^{-2k-1}\ol{\pp}^{k+1}\(\Omega^{k+1}\tilde{Y}\)\] \\
		&=\sum_{n=0}^{k+1}\binom{k+1}{n} \Omega^{k-1/2} \frac{(k+n)!}{k!}\(-\frac{\ii}{2L}\)^{n} \pp^{k+1}\[\Omega^{n-k}\]\ol{\pp}^{k+1-n}\tilde{Y} \\
		&=(k+1)\sum_{n=0}^{k+1}\frac{(k+n)!}{(n-k-1)!(k+1-n)!} \(-\frac{\ii}{2L}\)^{n}\(\frac{\ii}{2L}\)^{k+1} \Omega^{n+k+1/2}\ol{\pp}^{k+1-n}\tilde{Y} \\
		&= \frac{(k+1)(2k+1)!}{(4L^{2})^{k+1}} \Omega^{2k+3/2}\tilde{Y}(\zb)
	\end{aligned}
\end{equation}
and similarly
\begin{equation}
	\begin{aligned}
		\hat{\mathrm{O}}_{4} \tilde{Y}(\zb) &= \Omega^{k+1/2}\pp^{k}\[\Omega^{-2k-1}\ol{\pp}^{k}\(\Omega^{k+2}\tilde{Y}\)\] \\
		&= \sum_{n=0}^{k} \binom{k}{n} \frac{(k+1+n)!}{(k+1)!} \(- \frac{\ii}{2L}\)^{n} \Omega^{k+1/2} \pp^{k}\[\Omega^{n-k+1}\] \ol{\pp}^{k-n}\tilde{Y}(\zb)  \\
		&= \frac{1}{k+1} \sum_{n=0}^{k} \frac{(k+n+1)!}{(k-n)!(n-k)!} \(-\frac{\ii}{2L}\)^{n}\(\frac{\ii}{2L}\)^{k}\Omega^{n+k+3/2} \ol{\pp}^{k-n}\tilde{Y}(\zb) \\
		& = \frac{(2k+1)!}{(k+1)(2L)^{2k}}\Omega^{2k+3/2}\tilde{Y}(\zb).
	\end{aligned}
\end{equation}
By inspection
\begin{equation}
	\hat{\mathrm{O}}_{3}\tilde{Y}(\zb) = \(\frac{m_{k}}{2}\)^{2} \hat{\mathrm{O}}_{4} \tilde{Y}(\zb) ,
\end{equation}
and the second term in $\ol{\PP}\mathcal{F}_{z \dots z}\big|_{Y}$ vanishes.
Next, consider the holomorphic sector, for which we have
\begin{align}
    \Omega^{-k+1/2}\hat{\mathrm{O}}_{1}Y(z)&= \pp^{k+1}\[\Omega^{-2k-1}\pp^{k}\(\Omega^{k+2}Y(z)\)\] \nonumber \\
    &= \sum_{m=0}^{k}\binom{k}{m} \frac{(k+m+1)!}{(k+1)!} \(\frac{\ii}{2L}\)^{m}\pp^{k+1}\[\Omega^{m+1-k}\pp^{k-m}Y(z)\] \nonumber \\
    &= \sum_{m=0}^{k} \sum_{n=m}^{k+1+m} \binom{k}{m}  \binom{k+1}{n-m} \frac{(k+m+1)!}{(k+1)!} \(\frac{\ii}{2L}\)^{m}  \pp^{n-m}\Omega^{m+1-k}\pp^{2k+1-n}Y(z) \nonumber \\
    &= \sum_{n=k}^{2k+1} \frac{(2k+1)!}{(k+1-n+k)!}\(\frac{\ii}{2L}\)^{n}\Omega^{n+1-k}\pp^{2k+1-n}Y(z) \nonumber  \\
    & \quad + \sum_{n=0}^{k-1}(-)^{n}  \frac{k!}{(k-1-n)!} \(\frac{\ii}{2L}\)^{n} \Omega^{n+1-k}  \pp^{2k+1-n}Y(z) \[\sum_{m=0}^{n} (-)^{m} \binom{k+m+1}{n} \binom{n}{m} \frac{1}{k-m} \] \nonumber \\
    &= \sum_{n=0}^{2k+1} \frac{(2k+1)!}{(2k+1-n)!}\(\frac{\ii}{2L}\)^{n}\Omega^{n+1-k} \pp^{2k+1-n}Y(z),
\end{align}
and similarly
\begin{align}
    \Omega^{-k-1/2}\hat{\mathrm{O}}_{2}Y(z) &=  \pp^{k}\[\Omega^{-2k-1}\pp^{k+1}\(\Omega^{k+1}Y(z)\)\] \nonumber \\
    &= \sum_{m=0}^{k+1} \binom{k+1}{m}\frac{(m+k)!}{k!} \(\frac{\ii}{2L}\)^{m} \pp^{k}\[\Omega^{-k+m}\pp^{k+1-m}Y(z) \] \nonumber \\
    &= \sum_{m=0}^{k+1} \sum_{n=m}^{k+m} \binom{k+1}{m}\binom{k}{n-m}\frac{(k+m)!}{k!} \(\frac{\ii}{2L}\)^{m} \pp^{n-m}\Omega^{m-k}\pp^{2k+1-n}Y(z) \nonumber  \\
    &= \sum_{n=k+1}^{2k+1}\frac{(2k+1)!}{(2k+1-n)!} \(\frac{\ii}{2L}\)^{n}\Omega^{n-k}  \pp^{2k+1-n}Y(z) \nonumber \\
    &\quad + \sum_{n=0}^{k}(-)^{n}\frac{(k+1)!}{(k-n)!}\(\frac{\ii}{2L}\)^{n}\Omega^{n-k}\pp^{2k+1-n}Y(z)  \[\sum_{m=0}^{n} (-)^{m} \frac{1}{k+1-m}\binom{n}{m}\binom{k+m}{n}  \] \nonumber \\
    &= \sum_{n=0}^{2k+1}\frac{(2k+1)!}{(2k+1-n)!} \(\frac{\ii}{2L}\)^{n}\Omega^{n-k}  \pp^{2k+1-n}Y(z). 
\end{align}
Hence $\ol{\PP}\mathcal{F}_{z\cdots z}\big|_{Y} =0$ fully.

Next, consider the $\PP$ projection of the field strength
\begin{equation}
	\PP \mathcal{F}_{z\cdots z}\big|_{Y} = \nabla_{(-1/2)}^{k+1} \PP\Psi ,
\end{equation}
the only non-trivial component of which is
\begin{equation}
	\PP \mathcal{F}_{z\cdots z}\big|_{Y} \ni   \nabla_{(-1/2)}^{k+1}\[\Omega^{1/2}\mathcal{D}_{k+1}\big(\Omega^{-1}Y(z)\big)\] + \ii \frac{m_{k}}{2}\nabla_{(-1/2)}^{k+1}\[\Omega^{-1/2} \ol{\mathcal{D}}_{k}\big(\Omega \tilde{Y}(\zb)\big)\].
\end{equation}
Let's consider these terms separately.
First, we have
\begin{equation}
	\begin{aligned}
		\Omega^{-k-1/2}\nabla_{(-1/2)}^{k+1}&\[\Omega^{-1/2} \ol{\mathcal{D}}_{k}\big(\Omega \tilde{Y}(\zb)\big)\] = \pp^{k+1}\[\Omega^{-2k-2}\ol{\pp}^{k}\(\Omega^{k+2}\tilde{Y}(\zb)\)\] \\
		&=\sum_{n=0}^{k} \binom{k}{n} \pp^{k+1}\[\Omega^{-2k-2}\ol{\pp}^{n}\Omega^{k+2}\]\ol{\pp}^{k-n}\tilde{Y}(\zb) \\
		&= \sum_{n=0}^{k} \binom{k}{n} \frac{(k+1+n)!}{(k+1)!} \(-\frac{\ii}{2L}\)^{n} \pp^{k+1} \Omega^{n-k} \ol{\pp}^{k-n} \tilde{Y}(\zb) \\
		&= 0.
	\end{aligned}
\end{equation}
Next, 
\begin{align}
    \Omega^{-k-1/2}\nabla_{(-1/2)}^{k+1}&\[\Omega^{1/2}\mathcal{D}_{k+1}\big(\Omega^{-1}Y(z)\big)\] = \pp^{k+1}\[\Omega^{-2k-2}\pp^{k+1}\(\Omega^{k+1}Y(z)\)\] \nonumber \\
    &= \sum_{n=0}^{k+1}\binom{k+1}{n} \frac{(k+n)!}{k!} \(\frac{\ii}{2L}\)^{n} \pp^{k+1}\[\Omega^{n-k-1}\pp^{k+1-n}Y(z)\] \nonumber \\
    &= \sum_{n=0}^{k+1}\sum_{m=0}^{k+1} (-)^{n} \binom{k+1}{n}\binom{k+1}{m} \frac{(k+n)!}{k!} \frac{(k+1-n)!}{(k+1-n-m)!} \(\frac{\ii}{2L}\)^{m+n} \Omega^{m+n-k-1}  \pp^{2k+2-n-m}Y(z) \nonumber \\
    &= \sum_{m=0}^{k+1} \[\sum_{n=0}^{m} (-)^{n} \binom{k+1}{n} \binom{k+1}{m-n} \frac{(k+n)!}{k!} \frac{(k+1-n)!}{(k+1-m)!}\] \(\frac{\ii}{2L}\)^{m} \Omega^{m-k-1} \pp^{2(k+1)-m}Y(z) \nonumber \\
    &= \Omega^{-k-1} \pp^{2(k+1)}Y(z) + \sum_{m=1}^{k+1} \dots\,.
\end{align}
For $m \geq 1$, we can use Egorychev's method to evaluate the sum in square brackets in the ellipses:
\begin{equation}
	\begin{aligned}
		\sum_{n=0}^{m} (-)^{n} \binom{k+1}{n} \binom{k+1}{m-n} & \frac{(k+n)!}{k!} \frac{(k+1-n)!}{(k+1-m)!} =  \frac{k+1}{(m-1)!} \binom{k+1}{m} \sum_{n=0}^{m} (-)^{n} \binom{m}{n} \binom{k+n}{m-1} \\
		&= \frac{k+1}{(m-1)!} \binom{k+1}{m} \oint \frac{\dd u}{2\pi \ii}  \sum_{n=0}^{m} (-)^{n} \binom{m}{n} \frac{(1+u)^{k+n}}{u^{m}} \\
		&= \frac{k+1}{(m-1)!} \binom{k+1}{m} \oint \frac{\dd u}{2\pi \ii} \frac{(1+u)^{k}}{u^{m}} \[1-(1+u)\]^{m}   \\
		&= (-)^{m} \frac{k+1}{(m-1)!} \binom{k+1}{m} \oint \frac{\dd u}{2\pi \ii} (1+u)^{k} \\
		&= 0.
	\end{aligned}
\end{equation}
This shows that
\begin{equation}
	\PP \mathcal{F}_{z\cdots z}\big|_{Y} \ni \Omega^{-1/2}\pp^{2k+2}Y(z).
\end{equation}
Putting everything together, we arrive at \eqref{eq: field strength holomorphic splitting}.
The anti-holomorphic current follows similarly.

\section{Explicit invariance of the action \label{app:invariance of action}}

In this appendix, we show that the action \eqref{eq: dirac action} is explicitly invariant under the transformations \eqref{eq:covsymm}.

First note that the variation of the massive Majorana fermion action \eqref{eq: dirac action} is
\begin{equation}
    \delta_{\mathcal{L}}S=2(-1)^{|\mathcal{L}|}\int \dzz e \ol{\Psi}(\gamma^\mu\DD_\mu-\ii m_k\gamma_*)\delta_{\mathcal{L}}\Psi.
\end{equation}
Using \eqref{eq: Y main identity}, this can be simplified as
\begin{equation}
    \delta_{\mathcal{L}}S =2(-1)^{|\mathcal{L}|}\int \dzz e\ol\Psi \DD^{-}_{\mu_1}\cdots \DD^{-}_{\mu_{k+1}}\gamma_{\mu_{k+2}}\DD^{(\mu_1}\mathcal{Y}_{\mathcal{L}}^{\mu_2\cdots\mu_{k+2})_T},
\end{equation}
where the second term on the right-hand side of \eqref{eq: Y main identity} vanishes because $\mathcal{Y}_{\mathcal{L}}$ in \eqref{eq:covsymm} satisfies
\begin{equation}
    \gamma^{(\mu_1}\mathcal{Y}_{\mathcal{L}}^{\mu_2\cdots\mu_{k+2})_T}=0.
\end{equation}
Performing integration by parts $k+1$ times, the divergences become $\DD^{-}_{\mu_i}\mapsto -\DD^{+}_{\mu_i}$ acting on $\ol{\Psi}$.
Using \eqref{eq: field strength} and writing this in terms of the components of $\mathcal{F}$ and $\mathcal{Y}_{\mathcal{L}}$, we obtain
\begin{equation}
	\begin{aligned}
		\delta_{\mathcal{L}}S &=2(-1)^{|\mathcal{L}|+k+1}\int \dzz e\ol{\mathcal{F}}_{\mu_1\cdots\mu_{k+1}} \gamma_{\mu_{k+2}}\DD^{(\mu_1}\mathcal{Y}_{\mathcal{L}}^{\mu_2\cdots\mu_{k+2})_T} \\
		&= (-1)^{|\mathcal{L}|}\int \dzz F\ol\pp\mathcal{E}(\mathcal{L}).
	\end{aligned}
\end{equation}
Here, $F$ denotes $\Omega^{1/2}$ times the lower component of $\PP \mathcal{F}_{z\cdots z}$ off-shell. 
This integrand is a total derivative:
\begin{equation}
	\begin{aligned}
	    \delta_{\mathcal{L}}S&=(-1)^{|\mathcal{L}|}\int \dzz F\ol\pp\mathcal{E}(\mathcal{L})\\
	    &=-(-1)^{|\mathcal{L}|}\int \dzz \ol\pp F\sum_{i=0}^\infty(-1)^i\pp^i\frac{\pp}{\pp F_i}\mathcal{L}(\{F_n\},z)\\
	    &=-(-1)^{|\mathcal{L}|}\int \dzz \sum_{i=0}^\infty\ol\pp F_i\, \frac{\pp}{\pp F_i}\mathcal{L}(\{F_n\},z)\\
	    &=-(-1)^{|\mathcal{L}|}\int \dzz \ol\pp\mathcal{L}\\
	    &=0,
	\end{aligned}
\end{equation}
where in the second and third equality we used integration by parts, and in the fourth equality we used the chain rule. 
The action \eqref{eq: dirac action} is therefore invariant under \eqref{eq:covsymm}.

\end{document}